\documentclass[aps,nofootinbib,showpacs,preprintnumbers,amsmath,amssymb]{revtex4}
\def\be{\begin{equation}}
\def\ee{\end{equation}}
\def\ba{\begin{eqnarray}}
\def\ea{\end{eqnarray}}
\newcommand{\A}{{\mathcal{A}}}
\newcommand{\tA}{{\widetilde {\mathcal{A}}}}
\newcommand{\ta}{{\widetilde a}}
\newcommand{\tlA}{{\mathfrak A}}
\usepackage{graphics}
\usepackage{graphicx}
\usepackage{dcolumn}
\usepackage{bm}
\usepackage{epsfig}
\usepackage{graphicx}
\usepackage{multirow}
\usepackage{dcolumn}
\usepackage{graphicx,epsfig}%

\begin{document}

\preprint{arXiv:1006.5050v2, USM-TH-268, TUM-EFT-10/10}

\title{Simple analytic QCD model with perturbative QCD behavior at high momenta}

\author{Carlos Contreras$^1$}
\author{Gorazd Cveti\v{c}$^{1,}$$^2$}
\author{Olivier Espinosa\footnote{Deceased, September 14, 2010}$^1$}
\author{H{\'e}ctor E. Mart{\'\i}nez$^{1,}$$^3$}

\affiliation{$^1$Department of Physics, Universidad T{\'e}cnica Federico
Santa Mar{\'\i}a (UTFSM), Valpara{\'\i}so, Chile\\
$^2$Valparaiso Center for Science and Technology, UTFSM, Chile\\
$^3$Physik Department, Technische Universit\"at M\"unchen, D-85748 Garching, Germany}

\date{\today}

\begin{abstract}

Analytic QCD models are those where the QCD running coupling
has the physically correct analytic behavior, i.e., no Landau 
singularities in the Euclidean regime.
We present a simple analytic QCD model in which the discontinuity function
of the running coupling at high momentum scales is the same as in
perturbative QCD (just like in the analytic QCD model of Shirkov and
Solovtsov), but at low scales it is replaced by a delta function
which parametrizes the unknown behavior there.
We require that the running coupling agree to a high degree with the 
perturbative coupling at high energies, 
which reduces the number of free parameters of the model from 
four to one. The remaining parameter
is fixed by requiring the reproduction of the correct value of the 
semihadronic tau decay ratio.
\end{abstract}

\maketitle

\section{Introduction}
\label{sec:intr}

The running coupling $a_{\rm pt}(Q^2) \equiv \alpha_s(Q^2)/\pi$
in perturbative QCD (pQCD) possesses Landau
singularities at low momenta $Q^2$ ($0 < Q^2 < {\Lambda}^2$
where ${\Lambda}^2 \sim 0.1 \ {\rm GeV}^2$),
where $-q^2\equiv Q^2$ and $q$ being the typical momentum of
the considered process. These singularities
are then reflected in perturbative evaluation of spacelike physical
observables ${\cal D}(Q^2)$ in terms of powers of $a(\kappa Q^2)$
(with $\kappa \sim 1$).
On the other hand, in local field theories with causality
\cite{BS} the observables ${\cal D}(Q^2)$ are
analytic functions of $Q^2$ in the entire complex $Q^2$-plane,
with the exception of the negative semiaxis regime
$Q^2 \leq - M^2_{\rm thr}$, where the threshold mass $M_{\rm thr}
\sim 10^{-1}$ GeV reflects the threshold for production of (light) mesons.
If ${\cal D}(Q^2)$ is evaluated in terms of powers of $a(\mu^2)$
(with renormalization scale $\mu^2 = \kappa Q^2$, $\kappa = {\rm const.} \sim 1$), 
the aforementioned analytic behavior
of the true ${\cal D}(Q^2)$ should be reflected in $a(\kappa Q^2)$.
In perturbative QCD this is not the case, and the Landau singularities
of perturbative coupling $a_{\rm pt}(Q^2)$ should thus be regarded as unphysical.

On the other hand, analytic QCD models have
the running coupling $\A_1(Q^2)$ (instead of $a_{\rm pt}(Q^2)$)
which reflects more correctly
the analytic properties of ${\cal D}(Q^2)$.
Nonperturbative studies of ghost-gluon vertex and ghost and gluon propagators, 
using Schwinger-Dyson equations \cite{SDEs} and 
lattice calculations \cite{latt}, give QCD coupling $a(Q^2)$ 
with a finite value at $Q^2=0$ and without Landau singularities 
at positive $Q^2$.
The minimal analytic
(MA) model of Shirkov and Solovtsov \cite{ShS,MSS,Sh} is the first and
to date the most widely used of the analytic QCD models. It ``minimally''
modifies the perturbative coupling $a_{\rm pt}(Q^2)$,
by eliminating the offending part of the discontinuity
function of $a_{\rm pt}(Q^2)$ on the positive axis of $Q^2$-plane
($0 < Q^2 \leq {\Lambda}^2$) and keeping its discontinuity
on the nonpositive axis unchanged. This results in a new coupling
$\A_1^{\rm (MA)}(Q^2)$ analytic in the complex $Q^2$ plane with the
exception of the nonpositive semiaxis $-\infty < Q^2 \leq 0$. 
It turns out that, at high $Q^2$, the MA coupling differs from
the perturbative one by terms $\A_1^{\rm (MA)}(Q^2)-a_{\rm pt}(Q^2) \sim ({\Lambda}^2/Q^2)$.
Afterwards, other analytic models for $\A_1(Q^2)$ 
have been presented in the literature 
\cite{Nesterenko,Nesterenko2,Alekseev:2005he,Srivastava:2001ts,Webber:1998um,CV1,CV2}, 
and they satisfy certain 
additional constraints at low and/or at high $Q^2$.
For further literature and reviews of various analytic QCD models, 
see Refs.~\cite{Prosperi:2006hx,Shirkov:2006gv,Cvetic:2008bn}. 

Most of the analytic QCD models suffer from one or both of the
following problems:
\begin{enumerate}
\item
Analytic QCD models with few free parameters usually cannot reproduce
the experimental value of the semihadronic tau decay ratio $r_{\tau}$.
For example, MA is a model with this problem, 
cf.~Ref.~\cite{MSS,Milton:2000fi}. The
value of $r_{\tau}$ is at present the most precisely measured
low-energy QCD observable, with the squared momentum of the process
$s=|Q^2|=m^2_{\tau} \approx 3 \ {\rm GeV}^2$.
\item
The deviation of the analytic coupling $\A_1(Q^2)$ from the
perturbative coupling $a_{\rm pt}(Q^2)$, at high $Q^2 \gg {\Lambda}^2$,
can be appreciable: $\A_1(Q^2)-a_{\rm pt}(Q^2) \sim ({\Lambda}^2/Q^2)^k$
($k=1$, or $2$, ...). This implies that such analytic QCD model
gives nonperturbative contributions 
which are at least partly
from the ultraviolet (UV)
regime $\sim  ({\Lambda}^2/Q^2)^k$.
Such UV nonperturbative contributions contravene the
Operator product expansion (OPE) philosophy 
of the ITEP (Institute of Theoretical and Experimental Physics) 
group \cite{Shifman:1978bx},
which stipulates that the nonperturbative contributions
come only from the infrared (IR) regime.
\end{enumerate}

In Refs.~\cite{CV1,CV2} two analytic QCD models are constructed
such that the correct value of $r_{\tau}$ is reproduced;
however, at high $Q^2$ the deviations from perturbative QCD are
large, namely, $\A_1(Q^2)-a_{\rm pt}(Q^2) \sim ({\Lambda}^2/Q^2)^k$ with $k=1$.
Further, in  Refs.~\cite{CV1,CV2} a method is presented
for the construction of higher order analogs of powers $a_{\rm pt}^n(Q^2)$ 
for any analytic QCD model.

In Ref.~\cite{Alekseev:2005he} an analytic coupling $\A_1(Q^2)$ 
is constructed which comes close to fulfilling the
abovementioned ITEP-OPE condition, namely, it achieves
$\A_1(Q^2)-a_{\rm pt}(Q^2) \sim ({\Lambda}^2/Q^2)^k$ with $k=3$ at high $Q^2$.
However, the higher order analogs and 
the implications for the value of $r_{\tau}$ are not 
investigated in that reference.

In Refs.~\cite{CKV1,CKV2}, analytic coupling $\A_1(Q^2)$ is calculated
from certain classes of beta-functions, and such
$\A_1(Q^2)$ fulfills the OPE condition exactly,
i.e., $|\A_1(Q^2)-a_{\rm pt}(Q^2)| < ({\Lambda}^2/Q^2)^k$ for all $k>0$
at high $Q^2$. However, in such models it turns out
to be difficult to reproduce the correct value of $r_{\tau}$ 
\cite{CKV1,CKV2}.

In this paper we present a relatively simple analytic QCD model
which approximately fulfills
the OPE condition
(i.e., it merges with perturbative QCD to a high degree of accuracy
at high $|Q^2| > 10^1 \ {\rm GeV}^2$)
and, simultaneously, reproduces the correct value of $r_{\tau}$. 

In Section \ref{sec:mod} we present the model, by motivating
it first with a specific simple form of the discontinuity
function $\rho_1(\sigma)= {\rm Im} \A_1(Q^2=-\sigma- i \epsilon)$
in a specific renormalization scheme ($\beta_2=\beta_3=\cdots =0$).
In Section \ref{sec:res} we fix the free parameter by
requiring that the model reproduce the correct value of $r_{\tau}$.
There we also present results of the model for the Bjorken polarized 
sum rule (BjPSR) $d_{\rm Bj}(Q^2)$ at low $Q^2$ and compare them with experimental
values. In Section \ref{sec:concl} we present conclusions. 

\section{Model description}
\label{sec:mod}

The perturbative coupling $a_{\rm pt}(Q^2) \equiv \alpha_s(Q^2)/\pi$ 
has singularities (cut)
along the real $Q^2$ axis at $Q^2 \leq Q_b^2$ ($\sim {\Lambda}^2$) 
in the complex plane, where $Q_b^2$ is the branching point. 
Application of the Cauchy theorem
gives the following dispersion relation for $a_{\rm pt}(Q^2)$:
\be
a_{\rm pt}(Q^2) = \frac{1}{\pi} \int_{-Q_b^2 - \eta}^{+\infty} \ d \sigma 
\frac{ \rho_1^{\rm (pt)}(\sigma) }{(\sigma + Q^2)} \ ,
\label{disppt}
\ee
where $\rho_1^{\rm (pt)}(\sigma) = {\rm Im} \ a_{\rm pt}(Q^2=-\sigma - i \epsilon)$ is the perturbative discontinuity function, and $\eta \to +0$.
The minimal analytic (MA) model of Shirkov and Solovtsov \cite{ShS,MSS,Sh}
consists of removing the offending cut at positive $Q^2$
($0<Q^2 \leq Q_b^2$), i.e., for $-Q_b^2 \leq \sigma < 0$
\be
\A_1^{\rm (MA)}(Q^2) = \frac{1}{\pi} \int_0^{+\infty} \ d \sigma 
\frac{ \rho_1^{\rm (pt)}(\sigma) }{(\sigma + Q^2)} \ .
\label{dispMA}
\ee
In analogous way, the higher order couplings $\A_n$ ($n=2,3,\ldots$)
are constructed in MA analogously (\cite{MSS,Sh,Shirkov:2006gv} and
references therein)\footnote{
The MA couplings $\A_n^{\rm (MA)}$ ($n \geq 1$) defined here are the 
MA couplings of Refs.~\cite{ShS,Sh,Shirkov:2006gv} divided by $\pi$.}
\be
\A_n^{\rm (MA)}(Q^2) = \frac{1}{\pi} \int_0^{+\infty} \ d \sigma 
\frac{ \rho_n^{\rm (pt)}(\sigma) }{(\sigma + Q^2)} \ ,
\label{dispMAn}
\ee
where $\rho_n^{\rm (pt)}(\sigma) = {\rm Im} \ a_{\rm pt}^n (Q^2=-\sigma - i \epsilon)$. 

\begin{figure}[htb] 
\begin{minipage}[b]{.49\linewidth}
\centering\includegraphics[width=80mm]{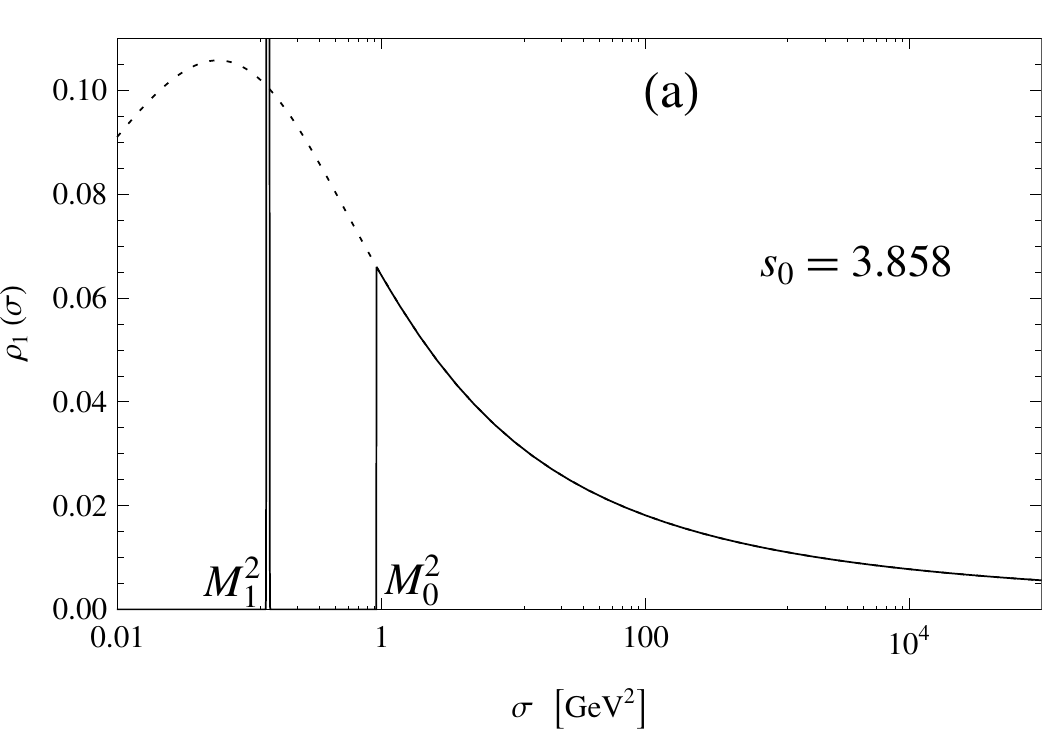}
\end{minipage}
\begin{minipage}[b]{.49\linewidth}
\centering\includegraphics[width=80mm]{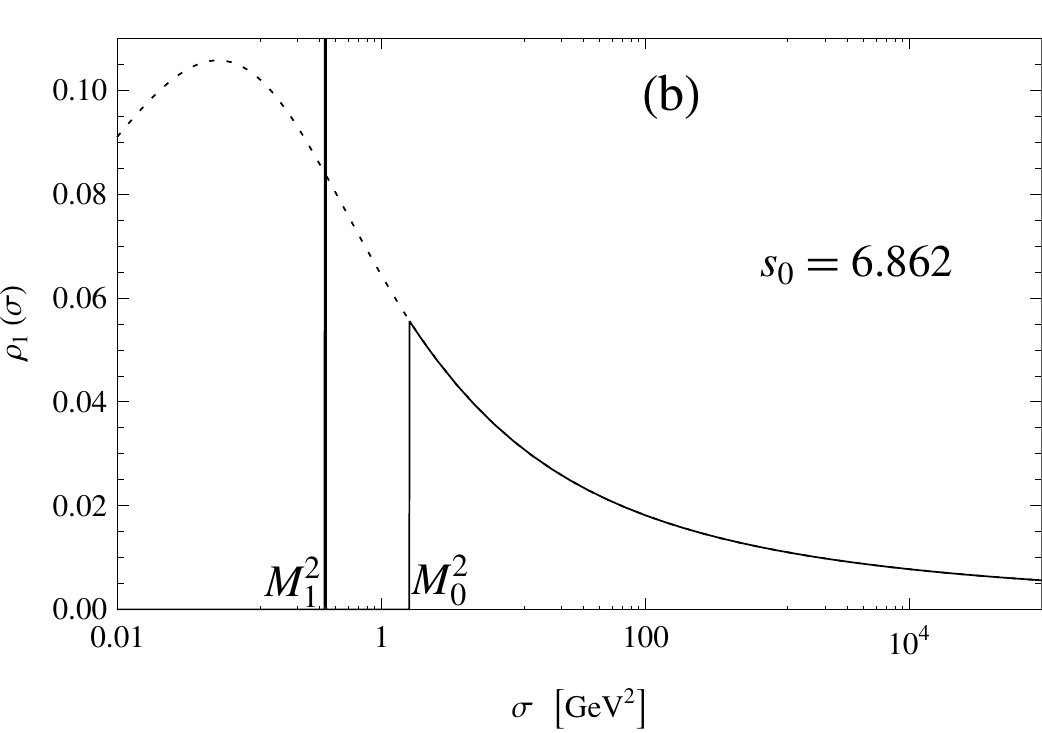}
\end{minipage}
\vspace{-0.4cm}
 \caption{\footnotesize The discontinuity function
$\rho_1(\sigma)$, Eq.~(\ref{rho1o1}), for the values of scale parameters
$\Lambda=0.487$ GeV and: (a) $M_1=0.371$ GeV and $M_0=0.956$ GeV 
($s_1 \equiv M_1^2/\Lambda^2 = 0.581$ and $s_0 \equiv M_0^2/\Lambda^2 = 3.858$); 
(b) $M_1=0.612$ GeV and $M_0=1.275$ GeV 
($s_1 \equiv M_1^2/\Lambda^2 = 1.579$ and $s_0 \equiv M_0^2/\Lambda^2 = 6.862$).
For comparison, perturbative discontinuity function $\rho_1^{\rm pt}(\sigma)$
at positive $\sigma$ is also included as a dotted curve
[in $\beta_2=\beta_3=\cdots=0$ renormalization scheme (RSch), and with $n_f=3$].}
\label{rho1mod}
 \end{figure}

As mentioned before, once the parameter ${\overline \Lambda}$ (the
${\overline {\rm MS}}$ $\Lambda$-scale) is adjusted in MA
so that the high-energy QCD phenomenology is reproduced (${\overline \Lambda}
\approx 400$-$440$ MeV when $n_f=3$, Ref.~\cite{Sh}), the value of the semihadronic
tau decay ratio $r_{\tau}$ (strangeless and massless) is predicted
to be about $0.14$, much lower \cite{Milton:2000fi,CV2} 
than the well measured experimental value $r_{\tau} \approx 0.203 \pm
0.004$. Another possibly unattractive aspect of MA is that its coupling 
$\A_1^{\rm (MA)}(Q^2)$ has singularities along the entire
nonpositive $Q^2$ axis ($Q^2 \leq 0$, i.e., $\sigma \geq 0$)
in the complex $Q^2$-plane. This does not reflect closely the
analyticity properties of spacelike observables 
${\cal D}(Q^2)$ which have nonanalyticity cut along
negative $Q^2$ axis starting at a negative threshold value 
$-M^2_{\rm thr}$: $Q^2 \leq - M^2_{\rm thr}$ (i.e.,
$\sigma \geq M^2_{\rm thr}$). The value of the threshold
mass $M_{\rm thr}$ is typically a (multiple of) mass of
light mesons. One possibility to incorporate such behavior
in the analytic coupling is to eliminate certain IR interval
$0 \leq \sigma < M_{\rm thr}^2$ of the cut in the dispersive
relation (\ref{dispMA}), resulting in a ``modified'' MA (mMA)
coupling
\be
\A_1^{\rm (mMA)}(Q^2) 
= \frac{1}{\pi} \int_{M_{\rm thr}^2}^{+\infty} \ d \sigma 
\frac{ \rho_1^{\rm (pt)}(\sigma) }{(\sigma + Q^2)} \ .
\label{dispA1mMA}
\ee
Such type of change was proposed in Refs.~\cite{Nesterenko2}.\footnote{
The authors of \cite{Nesterenko2} used for $\A_1$, before
eliminating an IR-cut, a ``second'' MA model \cite{Nesterenko},
which is obtained by minimal analytization of the right-hand side of
the following form of the perturbative renormalization group equation (RGE):
$d \ln a/d \ln Q^2 = -\beta_0 a - \beta_1 a^2 - \beta_2 a^2 - \cdots$.
Their numerical analysis is then performed at the one-loop level.}
In Ref.~\cite{CM}, it was pointed out that the coupling (\ref{dispA1mMA})
is a Stieltjes function, and that, as a consequence,
the paradiagonal Pad\'e approximants $[M-1/M](Q^2)$ of
such a coupling systematically converge to the exact values
$\A_1^{\rm (mMA)}(Q^2)$ when the Pad\'e index $M$ increases.\footnote{
Pad\'e approximant $[N/M](x)$ to a function $f(x)$ is, by definition,
a ratio $P_N(x)/Q_M(x)$ of polynomials of degree $N$ and $M$, respectively,
such that the Taylor expansions of $[N/M](x)$ and of $f(x)$ around
$x=0$ agree up to (and including) the term $\sim x^{N+M}$.}
The latter fact was checked numerically, and it was also shown there
that the aforementioned Pad\'e approximations are equivalent
to approximating the mMA discontinuity function by
a sum of $M$ delta terms
\ba
\frac{1}{\pi} \rho_1^{\rm (mMA)}(\sigma) &=&
\Theta(\sigma-M^2_{\rm thr}) \times {\rm Im} \ a_{\rm pt}(Q^2=-\sigma - i \epsilon)
\label{rho1mMA}
\\
& \approx & \sum_{n=1}^M f_n^2 \Lambda^2 \delta(\sigma - M_n^2) =
\sum_{n=1}^m f_n^2 \delta(s - s_n) \ ,
\label{rho1sumdel}
\ea
where $s=\sigma/\Lambda^2$, $s_n=M_n^2/\Lambda^2$ and $f_n$
are positive dimensionless quantities, and 
$M_{\rm thr} \approx M_1 < M_2 < \cdots$.
Therefore, it was argued that,
although a sum of delta functions appears to be a very crude
approximation for the (continuous) function $\rho_1(\sigma)$,
it gives increasingly better expressions for the coupling $\A_1(Q^2)$ when 
the number $M$ of deltas in the sum (\ref{rho1sumdel}) increases. 
We only know the approximate behavior of the true $\rho_1(\sigma)$ at high  
$\sigma \gg \Lambda^2$
[$\rho_1(\sigma) \approx \rho_1^{\rm (pt)}(\sigma)$], and we do not
know the behavior in the IR regime $\sigma \sim \Lambda^2$. 
Therefore, as was also argued in
Ref.~\cite{CM}, in the regime of low positive
$\sigma$ we can expect that parametrization of the true
$\rho_1(\sigma)$ in terms of one or a few delta functions
may lead to a reasonably realistic behavior of $\A_1(Q^2)$
at low $|Q^2|$. In this work we use only one delta function
in the IR regime of $\sigma$'s:
\ba
\rho_1(\sigma) &=&
\pi f_1^2 \Lambda^2 \; \delta(\sigma - M_1^2) +  \Theta(\sigma-M_0^2) \times 
\rho_1^{\rm (pt)}(\sigma) \ ,
\label{rho1o1}
\\
& = & \pi f_1^2 \; \delta(s - s_1) +  \Theta(s-s_0) \times r_1^{\rm (pt)}(s) \ ,
\label{rho1o2}
\ea
where $s=\sigma/\Lambda^2$, $s_1 = M_1^2/\Lambda^2$,
$s_0=M_0^2/\Lambda^2$, and $r_1^{\rm (pt)}(s) =  \rho_1^{\rm (pt)}(\sigma)
= {\rm Im} \ a_{\rm pt}(Q^2=-\sigma - i \epsilon)$.
We may expect\footnote{
Although we do not impose the condition $0 < M_1 < M_0$, it will turn out to be
true in our model (see the next Section).}
 $0 < M_1 < M_0$, i.e., the
actual threshold mass is $M_1$. The discontinuity function $\rho_1(\sigma)$,
Eq.~(\ref{rho1o1}), is depicted in Figs.~\ref{rho1mod} (a), (b), 
for the choice of two sets of values of the parameters 
($M_1$, $M_0$ and $\Lambda$), which will be motivated in the
next Section. 

A physical interpretation of the ansatz (\ref{rho1o1})-(\ref{rho1o2}) is
also possible: If the discontinuity function $\rho_1(\sigma)$ is to
simulate, in the first approximation, the spectral functions
$\rho_{\cal D}(\sigma) = {\rm Im} \ {\cal D}(Q^2=-\sigma - i \epsilon)$ of spacelike
observables ${\cal D}(Q^2)$, the delta-term in Eqs.~(\ref{rho1o1})-(\ref{rho1o2})
and in Figs.~\ref{rho1mod} can be regarded as a narrow width approximation (NWA) of
the dominant low-energy resonance.\footnote{
For two simpler-minded analytic QCD models involving delta-function in $\rho_1(\sigma)$, 
see Refs.~\cite{CV1,CV2}.} For an application of NWA ansatz directly to the spectral
function of the vector current-current correlator ($\Leftrightarrow$ Adler function),
see Ref.~\cite{Peris}. 

Furthermore, in a recent work of Ref.~\cite{Magradze:2010gn} a similar idea 
was applied directly to the spectral function $v_1(\sigma)$ 
of the vector current-current correlator.
The ansatz for $v_1(\sigma)$ there is similar to
the ansatz (\ref{rho1o1})-(\ref{rho1o2}). However, instead of approximating the
low-energy regime ($\sigma < M_0^2$) by a simple delta function, the known
measured low-energy values of $v_1(\sigma)$ of the ALEPH and OPAL collaborations
\cite{ALEPH1,ALEPH2,OPAL} were used. The threshold value of the onset of
perturbative QCD ($M_0^2$), and the
perturbative coupling strength ($\Leftrightarrow {\overline {\Lambda}}$) were then fixed by
two conditions: the correct measured value of the (strangeless) $r_{\tau}$ 
has to be reproduced,
and the higher-twist terms of the resulting Adler function have no terms of dimension two ($\propto 1/Q^2$). 
The work of Ref.~\cite{Magradze:2010gn} is a generalization and refinement of 
an earlier work Ref.~\cite{DeRafael}.

The dispersion relation applied to the discontinuity function
(\ref{rho1o1})-(\ref{rho1o2}) then gives us the following expression for the
analytic coupling $\A_1(Q^2)$: 
\ba
\A_1(Q^2) & = & \frac{1}{\pi} \int_{0}^{+\infty} d \sigma \; 
\frac{ \rho_1(\sigma) }{(\sigma + Q^2)} 
=  \frac{f_1^2}{u + s_1} + \frac{1}{\pi} \int_{s_0}^{\infty}
d s  \; \frac{ r_1^{\rm (pt)}(s) }{(s + u)} \ ,
\label{A1o2}
\ea
where we use the notation 
\be
u=Q^2/\Lambda^2, \qquad s=\sigma/\Lambda^2 \ .
\label{usnot}
\ee
In the renormalization scheme (RSch) with the beta function coefficients
$\beta_2=\beta_3=\cdots = 0$,
the discontinuity function is explicitly known in terms of Lambert
function $W_{\pm 1}$, Ref.~\cite{Gardi:1998qr} 
(see also Refs.~\cite{Magr1,Magr2} on the use of Lambert functions
in evaluation of the $n$-loop coupling)
\ba
r_1^{\rm (pt)}(s) & = & {\rm Im} \left[
\frac{1}{c_1} \frac{1}{\left[ 1 + W_{-1} \left( \frac{1}{c_1 e} s^{-\beta_0/c_1}
\exp [ - i \pi (\beta_0/c_1 - 1) ] \right) \right]} \right]
\nonumber\\
&=&  {\rm Im} \left[
\frac{-1}{c_1} \frac{1}{ \left[ 1 + W_{+1} \left( \frac{1}{c_1 e} s^{-\beta_0/c_1}
\exp[ + i \pi (\beta_0/c_1 - 1) ] \right) \right]} \right] \ ,
\label{r1tH}
\ea
where $s=\sigma/\Lambda^2$, and $\beta_0$ and $c_1=\beta_1/\beta_0$
are the two (universal) coefficients in the renormalization group equation
in the aforementioned renormalization scheme
\be
\frac{d a(Q^2)}{d \ln Q^2} = - \beta_0 a^2 (1 + c_1 a) \ ,
\label{RGEtH}
\ee
specifically $\beta_0=(11 - 2 n_f/3)/4$ and 
$\beta_1=\beta_0 c_1 = (102-38 n_f/3)/16$.
It is in this renormalization scheme that we will asume the form (\ref{rho1o2})
of the discontinuity function, and consequently,
the form (\ref{A1o2}) of $\A_1(Q^2)$.
The $\Lambda$ scale appearing implicitly in these expressions
is the Lambert $\Lambda$, and it is related with the
${\overline {\rm MS}}$ scale ${\overline \Lambda}$ at $n_f=3$ via
\be
\Lambda \approx {\overline \Lambda} \exp(0.3205) \ .
\label{LbL}
\ee
The coupling $a_{\rm pt}(Q^2)$ in the $\beta_2=\beta_3=\cdots = 0$
renormalization scheme has a Landau cut along $0 < Q^2 < Q_b^2$, where
$Q_b^2$ is the branching point $Q_b^2= \Lambda^2 s_b$,
where $s_b= c_1^{-c_1/\beta_0} \approx 0.635$ when $n_f=3$.
The perturbative coupling $a_{\rm pt}(Q^2)$
in this renormalization scheme is known \cite{Gardi:1998qr}
\be
a_{\rm pt}(Q^2) =
- \frac{1}{c_1} \frac{1}{\left[ 1 + W_{\mp 1} ( z_{\pm} ) \right]} \ ,
\label{apttH}
\ee
where $Q^2 = |Q^2| \exp( i \phi)$, the upper signs refer to the case
$0 \leq \phi < + \pi$, the lower signs to $- \pi < \phi < 0$, and
\be
z_{\pm} =  \frac{1}{c_1 e} \left( \frac{|Q^2|}{\Lambda^2} \right)^{-\beta_0/c_1} \exp \left[ i \left( \pm \pi - \frac{\beta_0}{c_1} \phi \right) \right] \ .
\label{zexpr}
\ee
It can be numerically checked that the dispersion relation
(\ref{disppt}) holds for $a_{\rm pt}(Q^2)$ in this renormalization scheme,
using for the discontinuity function the expressions
(\ref{r1tH}) at $s \equiv \sigma/\Lambda^2 >0$, and for
$-s_b < s \equiv \sigma/\Lambda^2 <0$ the expression
\ba
r_1^{\rm (pt)}(s) & = & {\rm Im} \left[
\frac{1}{c_1} \frac{1}{\left[ 1 + W_{-1} \left( \frac{-1}{c_1 e} |s|^{-\beta_0/c_1} + i \epsilon \right) \right]} \right] \ .
\label{r1sm0}
\ea
One peculiar feature in most of the analytic QCD models is that
at large $|Q^2|$ ($\gg \Lambda^2$) the analytic coupling
$\A_1(Q^2)$ differs from the perturbative coupling $a_{\rm pt}(Q^2)$
by an inverse power of $Q^2$:
\be
\vert \A_1(Q^2) - a_{\rm pt}(Q^2) \vert \sim \left( \frac{\Lambda^2}{Q^2}
\right)^k  \qquad  (|Q^2| \gg \Lambda^2) \ .
\label{devapt}
\ee
In MA \cite{ShS,MSS}, and in the ``second MA'' of Refs.~\cite{Nesterenko},
this power is $k=1$, i.e., relatively large deviation. 

However,
power deviations (\ref{devapt}) are not in accordance with the philosophy
of the Operator product expansion (OPE) as promoted by the 
ITEP group \cite{Shifman:1978bx}.
According to that philosophy, all nonperturbative contributions 
such as $(\Lambda^2/Q^2)^k$ to (inclusive) observables
originate from the infrared (IR) regimes $|Q^2| \alt \Lambda^2$. 
Consequently, the OPE of spacelike inclusive
observables is interpreted in this approach as superposition of perturbative
contributions coming from the ultraviolet regime $|Q^2| \gg \Lambda^2$
(Wilson coefficients) and nonperturbative contributions coming from the IR 
regime (vacuum expectation values of operators). Once we have power deviations
of the type (\ref{devapt}), this implies that a theory with such
$\A_1(Q^2)$ will give us, in evaluations of inclusive
spacelike observables, nonperturbative terms (of the type
$\Lambda^2/Q^2$) coming at least partly from the UV regime, 
thus contravening the ITEP philosophy.

More specifically, the authors of Ref.~\cite{DMW} argued 
in the following way that the terms $\sim (\Lambda^2/Q^2)^k$ 
in the deviation Eq.~(\ref{devapt}) contravene the ITEP philosophy of OPE.
Namely, let us consider the leading-$\beta_0$ summation of 
an inclusive spacelike observable ${\cal D}(Q^2)$
\be
{\cal D}^{\rm (LB)}(Q^2) \equiv 
\int_0^\infty \frac{dt}{t}\:  F_{\cal D}(t) \: 
a(t Q^2 e^{\overline {\cal C}}) \ .
\label{LBint}
\ee
Here, $F_{\cal D}(t)$ is the characteristic function of the
observable and ${\overline {\cal C}}=-5/3$. 
The quantity $t Q^2 e^{\overline {\cal C}}$ is the square of internal loop momenta
appearing in the resummation. In the UV regime $t > 1$, 
the deviation (\ref{devapt}) then 
produces power term contributions of UV origin in the observable
(see also Ref.~\cite{Cvetic:2007ad})
\be
\delta  {\cal D}^{\rm (LB)}(Q^2) \sim 
({\Lambda^2}/{Q^2} )^k \int_1^\infty \frac{dt}{t^{k+1}}\:  F_{\cal D}(t)
\sim ({\Lambda^2}/{Q^2} )^k \ .
\label{powerk2}
\ee
For more discussion on these aspects, we refer to Ref.~\cite{CKV2}.
The aforementioned OPE condition then implies
\be
\vert \A_1(Q^2) - a_{\rm pt}(Q^2) \vert < \left( \frac{\Lambda^2}{Q^2}
\right)^k  \qquad (|Q^2| \gg \Lambda^2; k=1,2,\ldots)
\label{ITEP}
\ee
This condition can be interpreted as a requirement that the
analytic QCD model be formally 
perturbative (analytic) QCD at high momenta. We want
to construct here a simple analytic QCD model with the discontinuity
function $\rho_1(\sigma)$ of the type (\ref{rho1o1}) which is
as close as possible to perturbative QCD 
(in renormalization scheme $\beta_2=\beta_3=\cdots = 0$).
In such a model, all the measured values of the high-energy QCD observables 
(with $|Q^2| > 10^1 \ {\rm GeV}^2$) are then reproduced just as in perturbative QCD. 
In such a model, we have to ensure only that the values of the well measured 
low-energy observables, particularly $r_{\tau}$ 
(with $|Q^2| = m_{\tau}^2 \approx 3 \ {\rm GeV}^2$), are reproduced.

It turns out to be difficult to construct analytic QCD theories which
fully respect the condition (\ref{ITEP}),
see Refs.~\cite{CKV1,CKV2}. We will
construct here an analytic QCD model of Eqs.~(\ref{rho1o1})-(\ref{A1o2}), 
which only approximately fulfills 
the condition (\ref{ITEP}). 
We will end up with a simple analytic QCD model where 
the number of free parameters is reduced to one,
and the latter parameter will be fixed by imposing the condition
of the reproduction of the correct value of $r_{\tau}$.
We note that the model of Eqs.~(\ref{rho1o1})-(\ref{A1o2})
at first contains three free dimensionless positive parameters
($f_1^2, s_1, s_0$), in addition to the energy scale
parameter $\Lambda$. 
The parameters $f_1^2, s_1$ and the scale
$\Lambda$ will be fixed by requiring: the condition that $k=3$ in the relation 
Eq.~(\ref{devapt}) instead\footnote{
This condition, with $k=3$, is also fulfilled in the model for $\A_1(Q^2)$
of Ref.~\cite{Alekseev:2005he}.}
of $k=1$; 
and the correct perturbative QCD value of $\A_1(Q^2)$ at $Q^2=(3 m_c)^2$.
This is approximately the highest value of $Q^2$ where
the number of active quark flavors can still reasonably safely be kept 
$n_f=3$ in perturbative QCD with four-loop RGE-running and three-loop 
matching conditions \cite{CKS}.\footnote{
If we apply the four-loop RGE-running in ${\rm {\overline MS}}$ renormalization scheme with
three-loop matching conditions according to Ref.~\cite{CKS}
at $Q^2 = (\kappa m_c)^2$ and $Q^2= (\kappa m_b)^2$, 
and choose a fixed initial value $a(m_c^2,{\overline {\rm MS}}; n_f=3)=0.12945$ 
at the initial $Q_0^2=m_c^2$ [which gives: $a( (3 m_c)^2,{\overline {\rm MS}};n_f=3)=0.07245$, 
i.e., the value we use in this paper], we obtain the values
$\alpha_s(M_Z^2,{\overline {\rm MS}})=0.1190, 0.1191, 0.1193$, when choosing
for the threshold parameter $\kappa$ the values $\kappa=3,2,1$, respectively.
For the quark masses we use the values $m_c=1.27$ GeV and $m_b=4.20$ GeV
\cite{PDG2008}.}

A consistent accounting of the threshold effects fully within analytic QCD, 
when $n_f \mapsto (n_f + 1)$, is not yet known for general analytic QCD models.
However, in MA of Refs.~\cite{ShS,MSS}, such an accounting can be made 
systematically in the following way 
(see Refs.~\cite{Sh}, and also Refs.~\cite{threshMA}): 
At the positive threshold value $Q^2_{\rm thr} = (\kappa m_c)^2$, with $\kappa \sim 1$ 
(usually in perturbative QCD: $1 \leq \kappa \leq 3$), 
the perturbative coupling can be taken to be continuous to a high degree of accuracy:
$a_{\rm pt}(Q^2_{\rm thr};n_f=3)=a_{\rm pt}(Q^2_{\rm thr};n_f=4)$.
The latter condition then relates the scale
parameters ${\overline \Lambda}(n_f=3)$ and ${\overline \Lambda}(n_f=4)$. In MA,
the discontinuity function is then constructed as
$\rho_1^{\rm (MA)}(\sigma) = {\rm Im} \ a_{\rm pt}(Q^2=-\sigma - i \epsilon; n_f)$ 
with $n_f=3$ for $\sigma < (\kappa m_c)^2$, and $n_f=4$ for
$(\kappa m_c)^2 < \sigma < (\kappa m_b)^2$.
Analogous threshold effects are implemented in $\rho_1^{\rm (MA)}$
for the transition $n_f=4 \mapsto 5$ at the threshold 
$\sigma_{\rm thr} = (\kappa m_b)^2$.
The function $\rho_1^{\rm (MA)}(\sigma)$ is thus step-like discontinuous.
The resulting (global) MA coupling $\A_1(Q^2)$, constructed via
the dispersion relation (\ref{dispMA}) with $\rho_1^{\rm (pt)}$ replaced by
the aforementioned step-wise discontinuous $\rho_1^{\rm (MA)}$,
remains analytic in the entire Euclidean region 
$Q^2 \in \mathbb{C} \backslash (-\infty, 0]$. The values of
${\overline \Lambda}(n_f)$'s can then be fixed by requiring that the
model reproduce the measured values of high-energy QCD observables
(e.g., with $|Q^2| > 10^1 \ {\rm GeV}^2$).

This threshold matching procedure could be implemented also in the present model
(in the renormalization scheme $\beta_2=\beta_3=\cdots=0$) by making $\rho_1(\sigma)$ accordingly
step-like discontinuous at $\sigma_{\rm thr}$'s. 
However, by assuming that at $Q^2_{\rm thr} = (3 m_c)^2$
the present $n_f=3$ analytic QCD model merges with perturbative QCD does not result in
any appreciable error, due to the condition that in the relation 
(\ref{devapt}) we have $k=3$ [see also the comments later in the text, the paragraph
just after Eq.~(\ref{incon}) and the paragraph at Eq.~(\ref{devapttAn})].

Throughout this work we assume that
we are in the regime of three active quark flavors ($n_f=3$),
and that the three quarks $u$, $d$ and $s$ are (almost) massless.
The ``approximate perturbative QCD'' condition $k=3$ in Eq.~(\ref{devapt})
can be expressed via the following two conditions:
\ba
\frac{1}{\pi} \int_{-s_b}^{+s_0} ds \; r_1^{\rm (pt)}(s) &=& f_1^2 \ ,
\label{ITEPk1}
\\
\frac{1}{\pi} \int_{-s_b}^{+s_0} ds \ s \; r_1^{\rm (pt)}(s) &=& s_1 f_1^2 \ ,
\label{ITEPk2}
\ea
where $s_b=c_1^{-c_1/\beta_0}=Q_b^2/\Lambda^2$, $Q_b^2$ being the
branch point of $a_{\rm pt}(Q^2)$ in the complex $Q^2$-plane.
Conditions (\ref{ITEPk1})-(\ref{ITEPk2}) mean that the coefficient
of $(Q^2/\Lambda^2)$  and $(Q^2/\Lambda^2)^2$
in the deviation $\A_1(Q^2)-a_{\rm pt}(Q^2)$ is zero, respectively.
Equations (\ref{ITEPk1})-(\ref{ITEPk2}) were obtained by subtracting Eq.~(\ref{A1o2})
from Eq.~(\ref{disppt})
\be
a_{\rm pt}(Q^2) - \A_1(Q^2) = - \frac{f_1^2}{u + s_1} + \frac{1}{\pi} \int_{-s_b}^{+s_0} 
d s  \; \frac{ r_1^{\rm (pt)}(s) }{(s + u)} \ ,
\label{difaA}
\ee
and expanding in powers of $(1/u) = (\Lambda^2/Q^2)$.

In addition to the two conditions (\ref{ITEPk1})-(\ref{ITEPk2}),
there is a condition that the theory merge with the
perturbative coupling (in the renormalization scheme $\beta_2=\beta_3=\cdots=0$) at
higher renormalization scales $\mu^2$. Since we do not
yet know a consistent exact threshold conditions within
analytic QCD models, we assume that our analytic QCD model coupling
$\A_1(\mu^2)$ has the number of active quarks $n_f=3$ up to the
renormalization scale (RScl)
$\mu^2=(3 m_c)^2$ (with $m_c \approx 1.27$ GeV) and that
at that scale it merges with the value of the perturbative coupling
$a_{\rm pt}((3 m_c)^2;\beta_2=\beta_3=\cdots=0;n_f=3)$
such as implied by the high energy QCD experiments.
Specifically, high energy QCD
implies $a(M_Z^2,{\overline {\rm MS}}) \approx 0.119/\pi$,
Ref.~\cite{PDG2008}. We then run this value, by
perturbative four-loop RGE in ${\overline {\rm MS}}$ renormalization scheme, 
down to renormalization scale $\mu^2= (3 m_c)^2$, and incorporate quark thresholds
at $\mu^2=(3 m_q)^2$ ($q=b,c$) by three-loop matching
conditions Ref.~\cite{CKS}. We obtain in this way 
${\overline a}_{\rm pt} \equiv 
a_{\rm pt}( (3 m_c)^2,{\overline {\rm MS}}, n_f=3) = 0.07245$. 
Changing renormalization scheme from ${\overline {\rm MS}}$ to
$\beta_2=\cdots=0$ by the subtracted form 
(Ref.~\cite{CK1}, Appendix A there) of the integrated perturbative QCD RGE 
(see Ref.~\cite{Stevenson}, Appendix A there) then results in\footnote{
We use for the ${\overline {\rm MS}}$ beta function $\beta(a(Q^2))$
at $Q^2=(3 m_c)^2$ and $n_f=3$ the Pad\'e ${\rm [2/3]}_{\beta}(a)$ based on the
4-loop polynomial  ${\overline {\rm MS}}$ beta function; 
if using the latter (polynomial) form,
we obtain slightly different value 
$a_{\rm in}$ [ $\equiv a_{\rm pt}((3 m_c)^2; \beta_2=0,\beta_3=0,\cdots; n_f=3)$] $=0.07054$.
It appears that, at such relatively large values of $a_{\rm pt}$, 
the Pad\'e ${\rm [2/3]}_{\beta}(a)$ is a better approximation to the full (yet unknown) 
${\overline {\rm MS}}$ beta function.} 
\be
a_{\rm in} \equiv a_{\rm pt}((3 m_c)^2; \beta_2=0,\beta_3=0,\ldots; n_f=3)
= 0.07050 \ .
\label{ain}
\ee
As stated above, we require that $\A_1(\mu^2)$ of our analytic QCD model, at
$\mu^2=(3 m_c)^2$, agrees with the value Eq.~(\ref{ain}),
i.e., the analytic QCD model merges with perturbative QCD starting at the
scale $\mu^2=(3 m_c)^2$ upwards
\be
\A_1(\mu^2=(3 m_c)^2) = a_{\rm in} \quad (= 0.07050) \ .
\label{incon}
\ee
One may worry that the replacement of the presented analytic QCD model
by perturbative QCD at $Q^2 \geq (3 m_c)^2$ 
[and with $n_f=3 \mapsto 4$ perturbative threshold at $Q^2 = (3 m_c)^2$]
may not be a good
approximation, i.e., that the analytic coupling $\A_1(Q^2)$ of the theory,
say at fixed $n_f=3$, behaves at high scales 
$Q^2 \sim M_Z^2$ significantly different than the perturbative coupling $a_{\rm pt}(Q^2)$.
It turns out that this is not the case. Namely, if we formally keep fixed 
$n_f=3$ (in order not to worry about threshold effects in analytic QCD), 
the perturbative RGE-running (in the renormalization scheme $\beta_2=\cdots=0$)
from the initial value $a_{\rm pt}((3 m_c)^2; n_f=3)=0.070502=0.221487/\pi$
[Eq.~(\ref{ain})] at the scale $Q^2=(3 m_c)^2$ to the high final scale
$Q^2=M_Z^2$ gives the value $a_{\rm pt}(M_Z^2; n_f=3)=
0.033694 = 0.105852/\pi$. The analytic coupling, for both
representative values of parameter $s_0$ used later in this work
($s_0=3.858, 6.862$), gives the same value $0.221487/\pi$ at $Q^2=(3 m_c)^2$, and
almost the same values at $Q^2=M_Z^2$: 
$\A_1(M_Z^2)=0.105853/\pi, 0.105856/\pi$, respectively.\footnote{ 
The values of $A_1(Q^2)$, for various values of the parameter $s_0$, differ from 
the values of the perturbative coupling only at low $Q^2 < 10 \ {\rm GeV}^2$;
for example, the relative difference $a_{\rm pt}(Q^2)/\A_1(Q^2) - 1$ is a monotonously
decreasing function of $Q^2$ for $Q^2 < 10 \ {\rm GeV}^2$; in
the interval $1 \ {\rm GeV}^2 < Q^2 < 5 \ {\rm GeV}^2$ it falls 
from $0.02$ to $0.0004$ when $s_0=3.858$, and from $0.05$ to $0.001$ when $s_0=6.862$.}
This strongly indicates that $\A_1(Q^2)$ of the analytic QCD model presented 
in this work, at both mentioned values of $s_0$, is practically indistinguishable 
from the perturbative $a_{\rm pt}(Q^2)$ at scales $Q^2 > (3 m_c)^2$. This conclusion
even gets generalized to any higher order couplings of this analytic QCD and
the perturbative QCD [see the comments in the paragraph at Eq.~(\ref{devapttAn})].

Altogether, the three conditions (\ref{ITEPk1}), (\ref{ITEPk2}),
and (\ref{incon}) eliminate three of the four otherwise
free parameters $s_0, s_1, f_1^2, \Lambda^2$ of our
analytic QCD model. We are thus left with only one free parameter,
e.g., the dimensionless parameter $s_0$ in Eq.~(\ref{A1o2})
for $\A_1(Q^2)$. In Table \ref{tabs1f1o2} we present
the numerical dependence of the parameters $s_1, f_1^2$ on $s_0$. 
It turns out that the value of the scale $\Lambda \approx 0.487$ GeV 
is practically independent of the value of $s_0$,
it varies by less than $0.1 \%$ for the range of the $s_0$-values
presented in Table \ref{tabs1f1o2}, the reason being that
$\A_1((3 m_c)^2) = a_{\rm pt}((3 m_c)^2)$ ($= 0.07050$)
behaves at such scales practically as perturbative coupling $a_{\rm pt}$
due to ``approximate perturbative QCD'' conditions (\ref{ITEPk1})-(\ref{ITEPk2}).\footnote{
For example, for the two input values $s_0=3.858, 6.862$ used later in the text,
we have $\Lambda=0.48679, 0.48687$ GeV, respectively; the perturbative value of $\Lambda$
(at $n_f=3$) is $\Lambda_{pQCD} =0.48676$ GeV.}
In the last two columns of Table \ref{tabs1f1o2} we include various
evaluated contributions to the strangeless and massless
semihadronic tau decay ratio $r_{\tau}$;
these aspects will be discussed in the
next section.\footnote{The value of $s_0=3.858$ results in the
reproduction of the central value of the experimental result 
$r_{\tau}(\Delta S=0, m_q=0)_{\rm exp}=0.203 \pm 0.004$ 
when the leading-$\beta_0$ (LB) and  
the beyond-the-leading-$\beta_0$ (bLB) contributions are
evaluated and added together.}
\begin{table}
\caption{The dimensionless nonnegative parameters $s_1=M_1^2/\Lambda^2$ 
and $f_1^2$ as functions of the cutoff parameter $s_0 = M_0^2/\Lambda^2$
($>0$). The scale $\Lambda$ is practically independent of $s_0$:
$\Lambda \approx 0.487$ GeV. The penultimate column shows the
leading-$\beta_0$ resummed (LB) contributions to $r_{\tau}$;
in parentheses the leading order (LO) contribution. The
last column shows the sum of the LB and the 
beyond-the-leading-$\beta_0$ contribution (LB+bLB) to $r_{\tau}$;
in parentheses the LO and beyond-leading-order contribution 
(LO+bLO) to $r_{\tau}$ (for details on $r_{\tau}$, 
see the next section).}
\label{tabs1f1o2}  
\begin{ruledtabular}
\begin{tabular}{lllll}
$s_0$ & $s_1$ & $f_1^2$ & $r_{\tau}^{\rm (LB)}$ ($r_{\tau}^{\rm (LO)}$) & $r_{\tau}^{\rm (LB+bLB)}$ ($r_{\tau}^{\rm (LO+bLO)}$)
\\ 
\hline
1.958 & 0.0000 & 0.1721 & 0.2522 (0.1315) & 0.2399 (0.1892)
\\
2.000 & 0.0121 & 0.1732 & 0.2509 (0.1315) & 0.2386 (0.1893)
\\
3.000 & 0.3117 & 0.1970 & 0.2290 (0.1316) & 0.2166 (0.1915)
\\
3.858 & 0.5812 & 0.2156 & 0.2156 (0.1317) & 0.2030 (0.1939)
\\
4.000 & 0.6267 & 0.2186 & 0.2137 (0.1317) & 0.2010 (0.1943)
\\
5.000 & 0.9523 & 0.2387 & 0.2016 (0.1319) & 0.1885 (0.1975)
\\
6.000 & 1.2861 & 0.2576 & 0.1916 (0.1321) & 0.1781 (0.2006)
\\
6.862 & 1.5788 & 0.2732 & 0.1844 (0.1323) & 0.1704 (0.2030)
\end{tabular}
\end{ruledtabular}
\end{table} 
In Table \ref{tabs1f1o2} we see that the cutoff parameter 
$s_0 = M_0^2/\Lambda^2$ cannot fall below $s_0 \approx 1.96$
because in such a case $s_1 = M_1^2/\Lambda^2$ turns out
to be negative and the coupling acquires a Landau singularity
(at $Q^2=-s_1 > 0$). 

\section{Evaluations of inclusive low-energy observables}
\label{sec:res}

In this Section we present the results of evaluation of
two inclusive low-energy observables in our discussed model.
We recall that the high-energy observables 
(with $|Q^2| \agt 10^1 \ {\rm GeV}^2$) are reproduced in
the model because at such energies the coupling
practically agrees with the perturbative coupling
\be
|\A_1(Q^2) - a_{\rm pt}(Q^2)| \alt (\Lambda^2/|Q^2|)^3  \quad
(|Q^2| \gg \Lambda^2) \ . 
\label{aprITEP}
\ee
The goal here is to fix the only free parameter $s_0$ of the model 
by requiring reproduction of the measured low energy QCD observables.
The most precisely measured inclusive low energy QCD observable
is the semihadronic tau decay ratio $R_{\tau}$,
which is the ratio of $\Gamma (\tau^- \to \nu_{\tau} {\rm hadrons} (\gamma) )$ and $\Gamma (\tau^- \to \nu_{\tau} e^- {\overline {\nu}_e} (\gamma))$.
After removing the (well measured) strangeness-changing
contribution, the color and Cabibbo-Kobayashi-Maskawa factors and
the electroweak effects, as well as the chirality-violating 
higher-twist (``quark mass'') contributions\footnote{
Further, assuming that the chirality-conserving higher twist effects are
negligible, i.e., that the gluon condensate
$\langle a G G \rangle$ is approximately zero.}
we obtain the experimental value
\be
r_{\tau}(\triangle S=0, m_q=0)_{\rm exp.} =
0.203 \pm 0.004 \ .
\label{rtauexp}
\ee
For details of this extraction we refer to Appendix B of Ref.~\cite{CKV2}
and references therein. The quantity $r_{\tau}$ of Eq.~(\ref{rtauexp})
is timelike, but it can be obtained from the
spacelike massless Adler function 
$d_{\rm Adl}(Q^2) =   a_{\rm pt}(Q^2) + {\cal O}(a_{\rm pt}^2)$ 
by contour integration \cite{Braaten:1988hc}
\be
r_{\tau} = \frac{1}{2 \pi} \int_{-\pi}^{+ \pi}
d \phi \ (1 + e^{i \phi})^3 (1 - e^{i \phi}) \
d_{\rm Adl} (Q^2=m_{\tau}^2 e^{i \phi}) \ ,
\label{rtaucont}
\ee
The perturbative expansion of $d_{\rm Adl}(Q^2)$
\be
d_{\rm Adl} (Q^2)_{\rm pt} = a_{\rm pt} + 
\sum_{n=1}^{\infty} (d_{\rm Adl})_n a_{\rm pt}^{n+1} \ .
\label{dAdlpt}
\ee
has been calculated up to ${\cal O}(a_{\rm pt}^4)$ \cite{d1,d2,d3},
i.e., the coefficients $(d_{\rm Adl})_n$ 
are known for $n=1,2,3$. Here, 
$a_{\rm pt} = a_{\rm pt}(\mu^2; c_2, c_3, \ldots)$ is
at a chosen renormalization scale $\mu^2$ and renormalization scheme ($c_2, c_3,\ldots$),
where $c_j = \beta_j/\beta_0$ ($j \geq 2$) are the renormalization scheme parameters.
For our purposes it is more convenient to reorganize the
expansion (\ref{dAdlpt}) in terms of the logarithmic derivatives
\be
\ta_{{\rm pt},n+1}(\mu^2)
\equiv \frac{(-1)^n}{\beta_0^n n!}
\frac{ \partial^n a_{\rm pt}(\mu^2)}{\partial (\ln \mu^2)^n} 
= a_{\rm pt}^n + {\cal O}(a_{\rm pt}^{n+1})
\qquad (n=1,2,\ldots) \ ,
\label{tan}
\ee
resulting in ``modified perturbation'' (mpt) series
\be
d_{\rm Adl}(Q^2)_{\rm mpt} = a_{\rm pt} 
+ \sum_{n=1}^{\infty} ({\widetilde d}_{\rm Adl})_n \ta_{{\rm pt},n+1} \ . 
\label{mpt}
\ee
The first three coefficients $({\widetilde d}_{\rm Adl})_n$
($n=1,2,3$) are known since they can be expressed via
$(d_{\rm Adl})_k$'s ($k=1,\ldots,n$). The basic idea of evaluation
of such leading-twist expressions in general analytic QCD models is
to replace 
(cf.~Refs.~\cite{CV1,CV2}; for MA, see also Ref.~\cite{Shirkov:2006gv})
\be
a_{\rm pt} \mapsto \A_1 \ , \quad
\ta_{{\rm pt},n+1} \mapsto \tA_{n+1}
\quad (n=1,2,\ldots) \ .
\label{anrule}
\ee
where $\tA_{n+1}$ are the corresponding logarithmic derivatives
in analytic QCD
\be
\tA_{n+1}(\mu^2) = \frac{(-1)^n}{\beta_0^n n!}
\frac{ \partial^n \A_1(\mu^2)}{\partial (\ln \mu^2)^n} \ ,
\qquad (n=1,2,\ldots) \ .
\label{tAn}
\ee
Therefore, the ``modified perturbation'' (mpt) series of perturbative QCD
(\ref{mpt}) is replaced in analytic QCD models by ``modified analytic''
(man) series
\be
d_{\rm Adl}(Q^2)_{\rm man} =
\A_1 + \sum_{n=1}^{\infty} ({\widetilde d}_{\rm Adl})_n \tA_{n+1} \ . 
\label{man}
\ee
The known truncated series of Adler function
in analytic QCD models is then
\be
d_{\rm Adl}(Q^2)_{\rm man}^{[N]} = \A_1 +
({\widetilde d}_{\rm Adl})_1 \tA_2 + 
\cdots 
({\widetilde d}_{\rm Adl})_{N-1} \tA_N \ ,
\label{manN}
\ee
with $N=4$, and with the coouplings $\A_1, \tA_2, \ldots$ at renormalization scale
\be
\mu^2 = Q^2 \exp({\cal C}) \quad ({\cal C} \sim 1) \ .
\label{RScl}
\ee
This series is then inserted into the contour integral
(\ref{rtaucont}), resulting in the sum
\be
r_{\tau}^{\rm (LO+bLO)[N]} = I(\A_1,{\cal C}) +
\sum_{n=1}^{N-1} \ ({\widetilde d}_{\rm Adl})_n I(\tA_{n+1},{\cal C}) \ ,
\label{rtLObLOmanN}
\ee
where $N=4$, and $I(\tA_{n+1},{\cal C})$ are the corresponding contour
integrals of $\A_1\equiv \tA_1$, $\tA_2, \ldots$
\be
I(\tA_{n+1},{\cal C}) =
\frac{1}{2 \pi} \int_{-\pi}^{+ \pi}
d \phi \ (1 + e^{i \phi})^3 (1 - e^{i \phi}) \
\tA_{n+1}(e^{\cal C} m_{\tau}^2 e^{i \phi}) \ ,
\label{IanC}
\ee 
The superscript ``LO+bLO'' in Eq.~(\ref{rtLObLOmanN}) indicates
that this is a sum of the leading order (LO) term
$I(\A_1,{\cal C})$ and of the higher order terms
beyond-the-leading-order (bLO).

The expansions (\ref{mpt}), (\ref{man})-(\ref{manN}) are 
expansions in nonpower quantities $\ta_{{\rm pt},n+1}$ and
$\tA_{n+1}$. The latter are constructed by applying logarithmic derivative
operators directly on the couplings $a_{\rm pt}$ and $\A_1$, and are thus
formally linear in the latter couplings. This is very convenient also for
the application of linear integral transforms on the observables, since
the linear integral transforms of $\A_1$ and $\tA_{n+1}$ usually become
simply related and since these transforms respect the truncation of the series.
In MA of Refs.~\cite{ShS,MSS}, the construction of
the higher couplings also has such properties, and therefore, the
transitions from momentum-transfer ($Q^2$) picture to the
energy ($s$ or $\sigma$) picture and to the distance ($r$) picture
become elegant and transparent, especially since the truncation of
such series is fully respected by the transformations 
(cf.~Refs.~\cite{Sh,threshMA}).

In this context, we stress that the spacelike observables
${\cal D}(Q^2)$, such as Adler function or Bjorken polarized sum rule
(BjPSR), at higher
momentum-transfer scales $Q^2 > (3 m_c)^2$ obtain practically
the same value when evaluated in the perturbative QCD by the truncated (modified)
perturbation series $d(Q^2)_{\rm mpt}^{[N]}$ [cf.~Eq.~(\ref{mpt})]
or evaluated in the presented analytic QCD model by the truncated
(modified) analytic series $d(Q^2)_{\rm man}^{[N]}$, Eq.~(\ref{manN}).
This is so because: 
\begin{itemize}
\item At such $Q^2$ the values of $a_{\rm pt}(Q^2)$
and $\A_1(Q^2)$ are practically equal, as a consequence of
the fulfilled condition (\ref{devapt}) with $k=3$ [see also the comments in
the paragraph just after Eq.~(\ref{incon})].
\item
Applying logarithmic derivatives to the relation (\ref{devapt}) with $k=3$,
valid in the presented analytic QCD model, we conclude that
\be
\vert \tA_{n+1}(Q^2) - \ta_{{\rm pt},n+1}(Q^2) \vert \sim 
\left( \frac{\Lambda^2}{Q^2}
\right)^3  \qquad  (|Q^2| \gg \Lambda^2; n=1,2,\ldots ) \ .
\label{devapttAn}
\ee
By analogy this implies that at $Q^2 > (3 m_c)^2$  the values of $\ta_{{\rm pt},n+1}$
(in the renormalization scheme $\beta_2=\beta_3= \cdots = 0$) 
and $\tA_{n+1}(Q^2)$ in the presented analytic QCD
are practically equal.
For example, if keeping $n_f=3$ fixed (in order not to worry about
the threshold effects), we obtain numerically: $\ta_2( (3 m_c)^2)=0.005593$ and 
$\tA_2((3 m_c)^2)=0.005592, 0.005587$ (when $s_0=3.858, 6.862$), 
and $\ta_2(M_Z^2)=0.0012033$ and $\tA_2(M_Z^2)=0.0012033,0.0012034$ 
(when $s_0=3.858, 6.862$); further, $\ta_3( (3 m_c)^2)=0.000468$ and
$\tA_3((3 m_c)^2)=0.000467, 0.000464$ (when $s_0=3.858, 6.862$), 
and $\ta_3(M_Z^2)=0.0000442 = \tA_3(M_Z^2)$ to the digits displayed 
(when $s_0=3.858, 6.862$).

\end{itemize}

There is yet another, more sophisticated, way of evaluating
inclusive spacelike and time-like observables. It is based on the
knowledge of the leading-$\beta_0$ part of coefficients
$d_n$ and ${\widetilde d}_n$ of the inclusive spacelike observable
such as $d_{\rm Adl}(Q^2)$. These leading-$\beta_0$ (LB) parts can then
be summed in any analytic QCD in the form
\be
(d_{\rm Adl})^{\rm (LB)}_{\rm an}(Q^2) \equiv 
\int_0^\infty \frac{dt}{t}\: F_{\rm Adl}^{\cal {E}}(t) \: 
\A_1(t Q^2 e^{\overline {\cal C}}) \ ,
\label{LB}
\ee
where ${\overline {\cal C}}=-5/3$. The Euclidean (${\cal {E}}$) characteristic 
function $F_{\rm Adl}^{\cal {E}}(t)$ is known \cite{Neubert}.
Expansion of expression (\ref{LB}) in logarithmic derivatives
$\tA_n(Q^2)$ then reproduces exactly the leading-$\beta_0$
part of the ``modified analytic'' expansion (\ref{man})
\be
(d_{\rm Adl})^{\rm (LB)}_{\rm an}(Q^2) =
\A_1 + \sum_{n=1}^{\infty} c^{(1)}_{nn} \beta_0^n
\tA_{n+1} \ .
\label{LBexp}
\ee
where the expansion of each coefficient $({\widetilde d}_{\rm Adl})_n$
and $(d_{\rm Adl})_n$ of the ``mpt'' and ``pt'' expansions,
Eqs.~(\ref{mpt}) and (\ref{dAdlpt}), in powers of $\beta_0$ is
\be
({\widetilde d}_{\rm Adl})_n = 
\sum_{k=-1}^{n} {\widetilde c}^{(1)}_{nk} \beta_0^k \ ,
\quad
(d_{\rm Adl})_n = 
\sum_{k=-1}^{n} c^{(1)}_{nk} \beta_0^k \ , 
\quad
{\widetilde c}^{(1)}_{nn} = c^{(1)}_{nn} \ .
\label{tdns}
\ee
In practice, we know the full coefficients $({\widetilde d}_{\rm Adl})_n$
for $n=1,2,3$. Subtracting from them the leading-$\beta_0$
parts $c^{(1)}_{nn} \beta_0^n$ then gives us a truncated series
for the beyond-the-leading-$\beta_0$ (bLB) contributions.
Hence, the LB-resummed expression (LB+bLB) for the Adler function, in any
analytic QCD model, is 
\ba
(d_{\rm Adl})_{\rm man}^{\rm (LB+bLB)}(Q^2)^{[N]} & = &
\int_0^\infty \frac{dt}{t}\: F_{\rm Adl}^{\cal {E}}(t) \: 
\A_1(t Q^2 e^{\overline {\cal C}}) +
\sum_{n=1}^{N-1} (T_{\rm Adl})_n \tA_{n+1} \ ,
\label{LBbLBmanN}
\ea
where $N=4$ and
\be
(T_{\rm Adl})_n = ({\widetilde d}_{\rm Adl})_n - c^{(1)}_{nn} \beta_0^n \ .
\label{Tn}
\ee
When inserting the expression (\ref{LBbLBmanN}) into the
contour integral (\ref{rtaucont}), we get the LB-resummed
expression for $r_{\tau}$ in any analytic QCD model
\be
r_{\tau}^{\rm (LB+bLB),[N]} = r_{\tau}^{\rm (LB)} + 
\sum_{n=1}^{N-1} \ (T_{\rm Adl})_n I(\tA_{n+1},{\cal C}) \ ,
\label{rtLBbLBmanN}
\ee
where $I(\tA_{n+1},{\cal C})$ are given in Eq.~(\ref{IanC}),
and the LB-part is obtained by using the LB-integral (\ref{LB})
in contour integral (\ref{rtaucont}). In Ref.~\cite{Neubert2}
this expression was expressed in terms of the Minkowskian
coupling $\tlA_1(s)$
\ba
\tlA_1(s) &=& \frac{1}{\pi} 
\int_s^{\infty} \frac{d \sigma}{\sigma}
{\rho}_1(\sigma) \ ;
\label{tlA1}
\ea
in the form 
\be
r_{\tau}^{\rm (LB)} = 
\int_0^\infty \frac{dt}{t}\: F_{r}^{\cal {M}}(t) \: 
\tlA_1 (t e^{\cal C} m_{\tau}^2) \ .
\label{LBrt}
\ee
where the Minkowskian characteristic function $F_{r}^{\cal {M}}(t)$
was calculated explicitly.\footnote{Since we use a different normalization,
$F_{r}^{\cal {M}}(t)$ here is equal to $(t/4)$ times $F_{r}^{\cal {M}}(t)$ of
Ref.~\cite{Neubert2}.} Using the relation $d \tlA_1(s)/d \ln s = -\rho_1(s)/\pi$ and
performing integration by parts, the LB-contribution to (the
massless and strangeless) $r_{\tau}$ can be rewritten in the following
form in terms of the discontinuity function $\rho_1(\sigma)$, 
which is more convenient for numerical evaluations:
\be
r_{\tau}^{\rm (LB)} = 
\frac{1}{\pi} \int_0^\infty \frac{dt}{t}\: {\widetilde F}_{r}(t) \: 
\rho_1(t e^{\cal {\overline C}} m_{\tau}^2) \ ,
\label{LBrt2}
\ee
where 
\be
{\widetilde F}_{r}(t) = \int_0^t \frac{dt'}{t'}\: F_{r}^{\cal {M}}(t') \ .
\label{tFtau}
\ee 
This expression was used in Refs.~\cite{CKV1,CKV2}, and explicit expression 
for the characteristic function ${\widetilde F}_{r}(t)$ is given in
Ref.~\cite{CKV2} (Appendix D).

For derivations of and more details on the above identities, we
refer to Refs.~\cite{CV1,CV2,CKV1,CKV2}.

For the evaluations of the (strangeless and massless) $r_{\tau}$
in our model, we use the LB-resummed expression (\ref{rtLBbLBmanN})
and, as an alternative, the more rudimentary expression
(\ref{rtLObLOmanN}), both with $N=4$.
In Table \ref{tabs1f1o2}, of the previous Section, we included the
results of these evaluations in the last two columns. We used
the renormalization scale parameter ${\cal C}=0$, i.e., $|\mu^2| = m_{\tau}^2$ on the
contour integral [cf.~Eqs.~(\ref{RScl}), (\ref{IanC})]. We note
that the leading-$\beta_0$ contribution Eq.~(\ref{LBrt2}) [and: Eq.~(\ref{LB})]
is independent of renormalization scale. From the results of Table \ref{tabs1f1o2}
we see that, when the value of the model parameter $s_0=M_0^2/\Lambda^2$ 
increases, the value of $r_{\tau}$ slowly decreases in the
LB-resummed approach (LB+bLB) Eq.~(\ref{rtLBbLBmanN}), 
and slowly increases in the other (LO+bLO) approach Eq.~(\ref{rtLObLOmanN}). 
The approach of Eq.~(\ref{rtLBbLBmanN}) reproduces the central 
experimental value (\ref{rtauexp}) $r_{\tau}=0.203$ at the parameter value 
$s_0=3.858$; the approach of Eq.~(\ref{rtLObLOmanN})
reproduces $r_{\tau}=0.203$ at $s_0=6.862$. 

\begin{figure}[htb] 
\begin{minipage}[b]{.49\linewidth}
\centering\includegraphics[width=80mm]{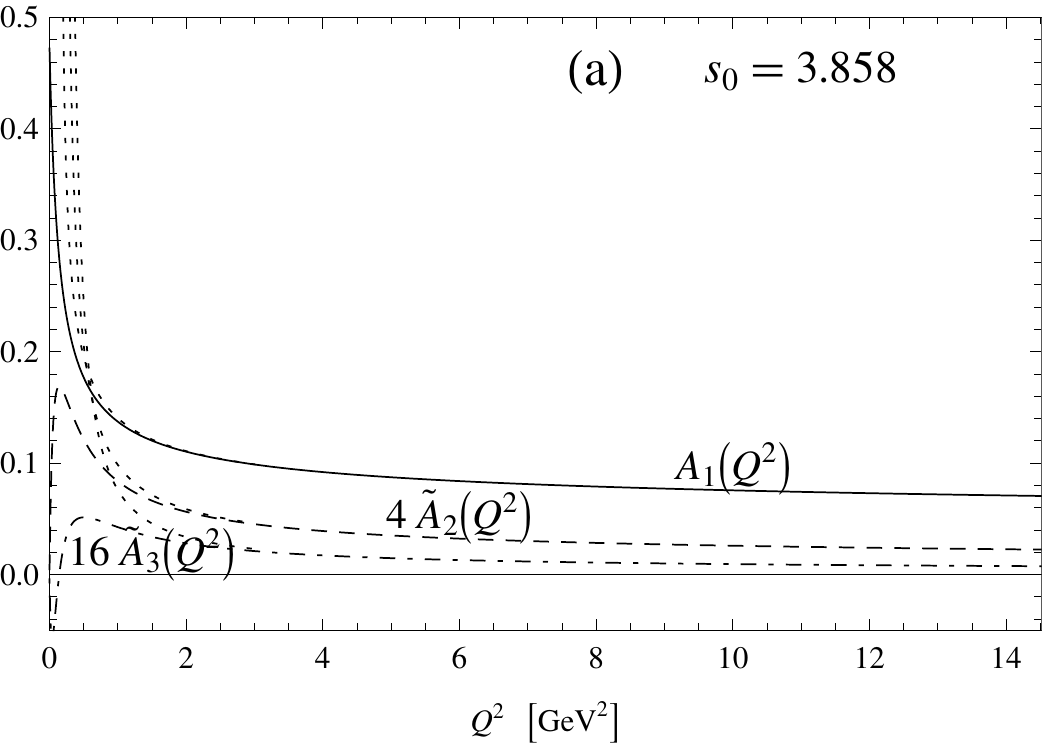}
\end{minipage}
\begin{minipage}[b]{.49\linewidth}
\centering\includegraphics[width=80mm]{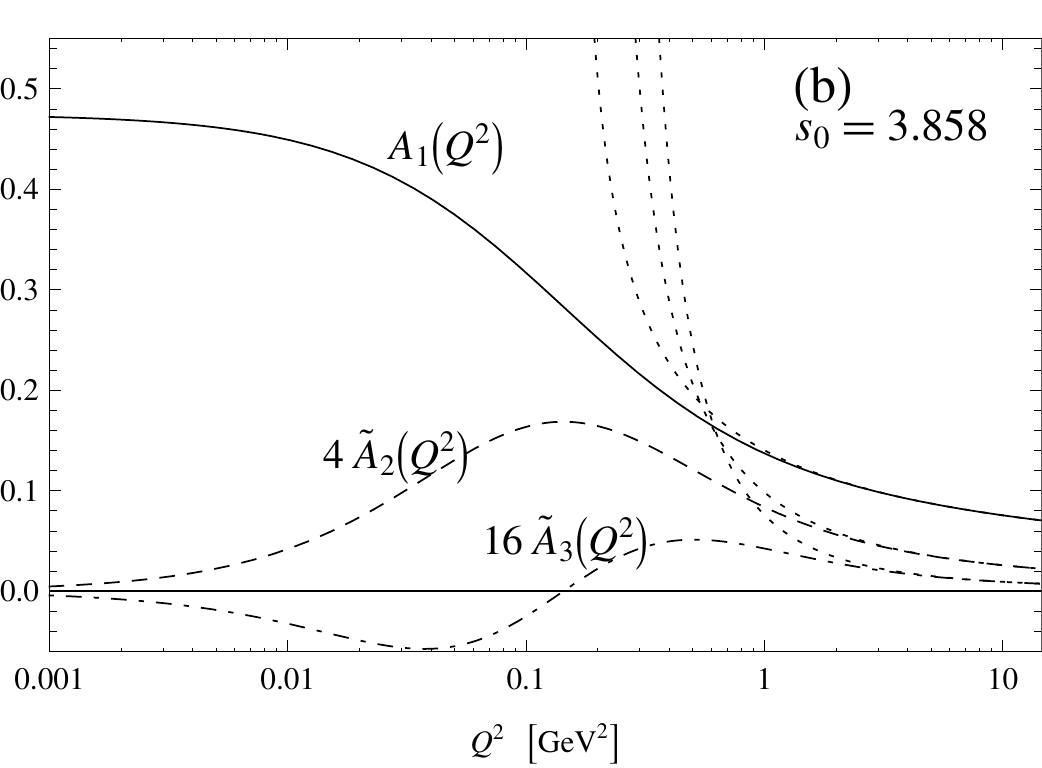}
\end{minipage}
\vspace{-0.4cm}
 \caption{\footnotesize The analytic couplings $\A_1(Q^2)$ (full curve),
$\tA_2(Q^2)$ (dotted curve) and $\tA_3(Q^2)$ (dot-dashed curve)
for positive $Q^2$ ($0 \leq Q^2 \leq (3 m_c)^2$) for the choice of parameter
$s_0= 3.858$ when: (a) linear scale is used for $Q^2$; (b) logarithmic scale 
is used for $Q^2$. For comparison, the corresponding perturbative couplings
$a_{\rm pt}(Q^2)$, ${\tilde a}_{{\rm pt},2}(Q^2)$ and 
${\tilde a}_{{\rm pt},3}(Q^2)$ are included as dotted curves
(in the renormalization scheme $\beta_2=\beta_3=\cdots=0$, with $n_f=3$). The couplings
$\tA_2(Q^2)$ and ${\tilde a}_{{\rm pt},2}(Q^2)$ were rescaled by factor 4,
and the couplings $\tA_3(Q^2)$ and ${\tilde a}_{{\rm pt},3}(Q^2)$ by factor 16,
for better visibility.}
\label{Asoa}
 \end{figure}

\begin{figure}[htb] 
\begin{minipage}[b]{.49\linewidth}
\centering\includegraphics[width=80mm]{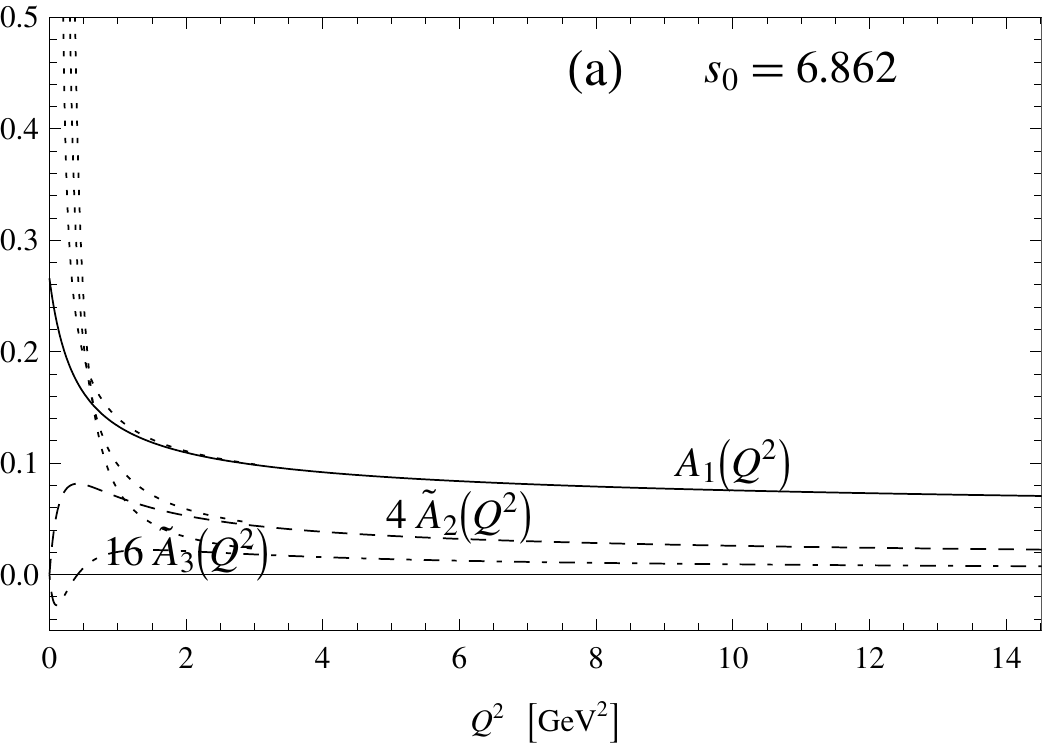}
\end{minipage}
\begin{minipage}[b]{.49\linewidth}
\centering\includegraphics[width=80mm]{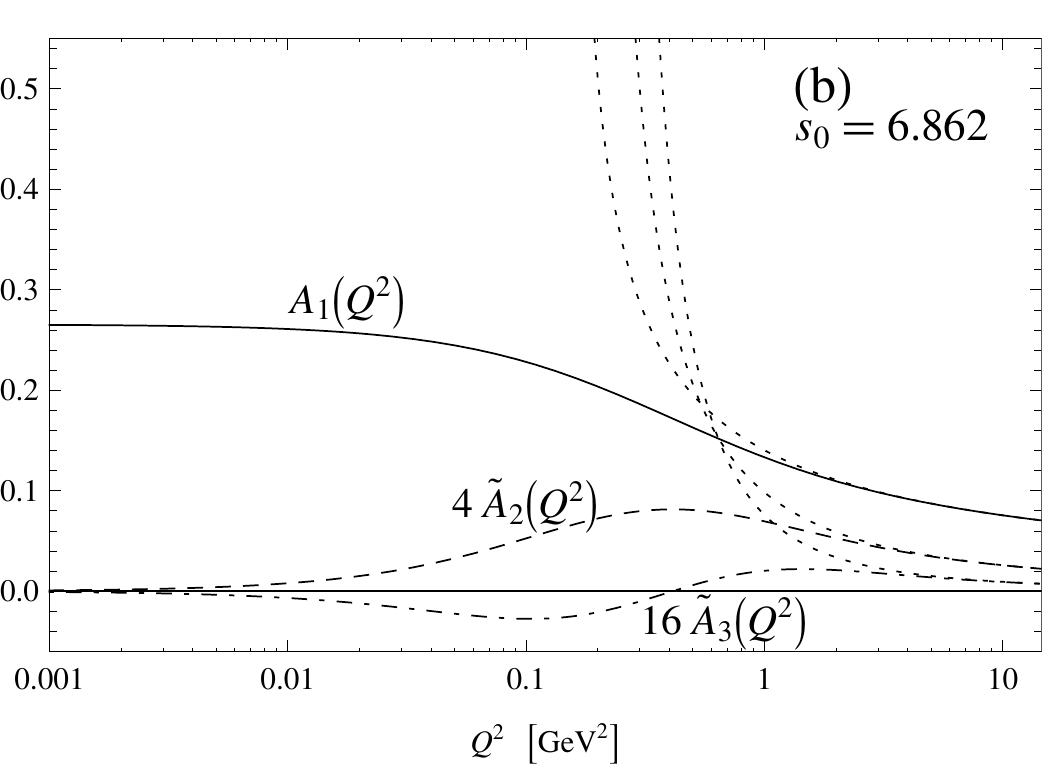}
\end{minipage}
\vspace{-0.4cm}
 \caption{\footnotesize Same as in Figs.~\ref{Asoa} (a), (b), but now
for the choice of the analytic QCD parameter $s_0=6.862$.}
\label{Asob}
 \end{figure}

The running couplings $\A_1(Q^2)$ and $\tA_n(Q^2)$ ($n=2,3$)
for positive $Q^2$ ($0 \leq Q^2 \leq (3 m_c)^2$) are depicted in 
Figs.~\ref{Asoa} and \ref{Asob} for the aforementioned choices of parameter
values ($s_0= 3.858$ and $6.862$), respectively. In these Figures, the 
perturbative couplings (in the same $\beta_2=\beta_3=\cdots=0$ renormalization scheme) are also
presented, for comparison. We note that the analytic coupling
$\A_1(Q^2)$ coincides with the perturbative coupling $a_{\rm pt}(Q^2)$
at $Q^2 = (3 m_c)^2$. The conditions (\ref{ITEPk1})-(\ref{ITEPk2})
bring the behavior of the analytic couplings very close to those of
perturbative couplings in the high momentum regime $|Q^2| \gg \Lambda^2$
(note: $\Lambda^2 \approx 0.24 \ {\rm GeV}^2$). This is clearly seen
in Figs.~\ref{Asoa} and \ref{Asob} where the perturbative
couplings $a_{\rm pt}$, ${\tilde a}_{{\rm pt},2}$ and 
${\tilde a}_{{\rm pt},3}$ (dotted curves)
virtually agree with the corresponding analytic couplings
$\A_1$, $\tA_2$ and $\tA_3$ for $Q^2$ down to $Q^2 \approx 2 \ {\rm GeV}^2$.
The Landau singularities of the perturbative couplings appear for 
$Q^2 \leq Q_b^2$ where the branching point is $Q_b^2= c_1^{-c1/\beta_0} \Lambda^2$ 
($\approx 0.635 \Lambda^2 \approx  0.1504 \ {\rm GeV}^2$ when $n_f=3$), and perturbative
couplings diverge at the branching point $Q^2=Q_b^2$ (Landau pole).
The discontinuity function
$\rho_1(\sigma) = {\rm Im} \A_1(Q^2=-\sigma - i \epsilon)$ for
the two cases $s_0= 3.858$ and $6.862$ is presented in Figs.~\ref{rho1mod}
(a), (b), respectively; it differs very much from the
perturbative analog $\rho_1^{\rm pt}(\sigma)$ which is also included there.

\begin{figure}[htb] 
\centering\includegraphics[width=120mm]{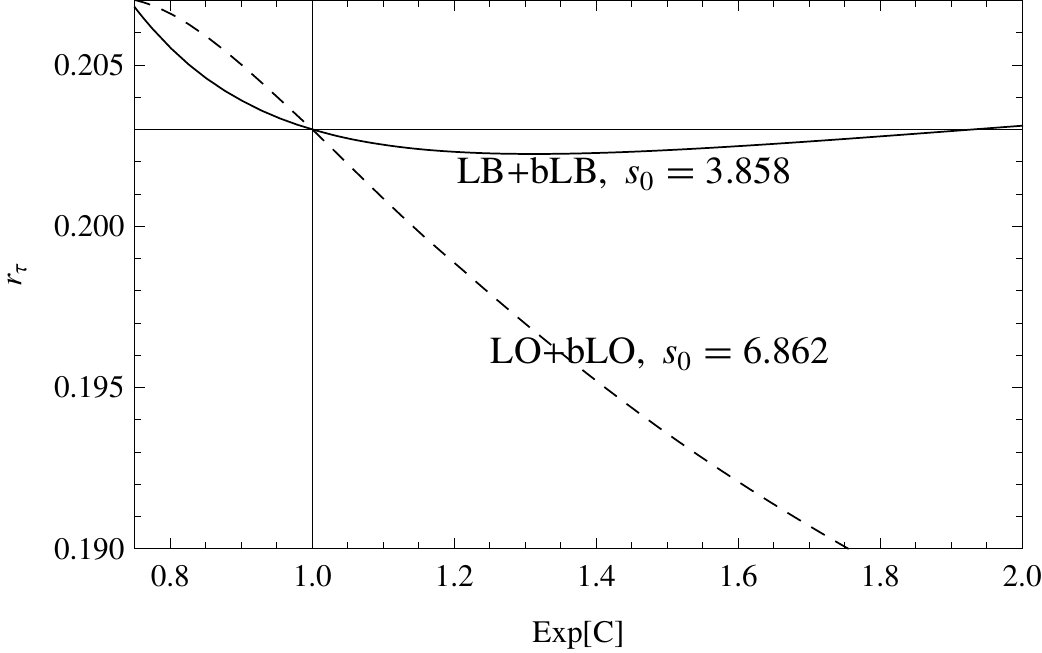}
\vspace{-0.4cm}
 \caption{\footnotesize The dependence of the evaluation
method of Eq.~(\ref{rtLBbLBmanN}) (LB+bLB) and
of Eq.~(\ref{rtLObLOmanN}) (LO+bLO), for $r_{\tau}$, 
of the contour renormalization scale parameter 
$\exp({\cal C}) \equiv |\mu^2|/m_{\tau}^2$.
Values of the $s_0$ parameter ($3.858$ and
$6.862$, respectively) were adjusted so that,
at $\exp({\cal C})=1$, the central experimental value
$r_{\tau}(\Delta S=0,m_q=0) = 0.203$ is reproduced.}
\label{rtauvsRScl}
 \end{figure}

A closer look at the two approaches, Eqs.~(\ref{rtLBbLBmanN}) and
(\ref{rtLObLOmanN}), reveals:
\begin{itemize}
\item 
The simple approach Eq.~(\ref{rtLObLOmanN}) 
for $r_{\tau}$ in this model requires 
quite a large IR cut $M_0$ for the MA-discontinuity function
$M_0 = \Lambda \sqrt{s_0} \approx 1.28$ GeV which appears to be
dangerously close to the mass of the process $m_{\tau}=1.777$ GeV;
in such a case the scales $|Q| \approx m_{\tau}$ are close to the
energy regime $0 < \sigma < M_0$ where the discontinuity function
$\rho_1(\sigma)$ is parametrized by only one delta function --
cf.~Fig.~\ref{rho1mod}(b).
On the other hand, the LB-resummed approach Eq.~(\ref{rtLBbLBmanN})
requires that the IR cutoff be $M_0 = \Lambda \sqrt{s_0} \approx  0.96$ GeV,
roughly half of the mass $m_{\tau}$ -- cf.~Fig.~\ref{rho1mod}(a).
\item
The convergence properties of the truncated ``modified analytic''
sum Eq.~(\ref{rtLObLOmanN}) for $r_{\tau}$ show that
the last (fourth) term is appreciable ($\approx 0.017$). 
On the other hand, the last (fourth) term in the sum Eq.~(\ref{rtLBbLBmanN}) 
is significantly smaller ($\approx 0.005$) -- see Table \ref{tabrt} in
the case of the renormalization scale parameter ${\cal C}=0$. 
\item
When varying the parameter ${\cal C}$, Eq.~(\ref{RScl}), 
away from ${\cal C}=0$ upwards, e.g., in the interval between $0$
and $\ln(2)$ (i.e., $|\mu^2|$ on the contour between $m_{\tau}^2$
and $2 m_{\tau}^2$), the result for $r_{\tau}$ in the approach of
Eq.~(\ref{rtLObLOmanN}) decreases by several percent, while in the 
leading-$\beta_0$ resummed approach of Eq.~(\ref{rtLBbLBmanN}) 
it remains virtually unchanged -- see Table \ref{tabrt}.
When moving ${\cal C}$ to negative values ($|\mu^2| < m_{\tau}^2$),
the two approaches have mutually comparable stronger renormalization scale dependence,
something to be expected since the model is apparently a
simple approximation to the true situation for $\rho_1(\sigma)$
at low $\sigma < m_{\tau}^2$.
\end{itemize} 
\begin{table}
\caption{The four terms in truncated analytic expansions
(\ref{rtLBbLBmanN}) and (\ref{rtLObLOmanN}) for $r_{\tau}$,
for the values of the $s_0$ parameter $s_0=3.858$ and $s_0=6.862$,
respectively. The renormalization scale parameter ${\cal C}$ 
is varied from $\ln(3/4)$
to $\ln 2$. For ${\cal C}=0$, the two methods, with
their respective values of $s_0$, reproduce the
central experimental value $r_{\tau}=0.203.$}
\label{tabrt}  
\begin{ruledtabular}
\begin{tabular}{ll|llll|l}
method, $s_0$ & ${\cal C}$ & $r_{\tau}:$ LB (LO) & NLB (NLO) & 
${\rm N}^2{\rm LB}$ (${\rm N}^2{\rm LO}$) &
${\rm N}^3{\rm LB}$ (${\rm N}^3{\rm LO}$) &
sum (sum) 
\\
\hline
\multirow{4}{15mm}{LB+bLB, $s_0=3.858$} & 
$\ln(0.75)$   & 0.2156 & 0.0018 & -0.0258 & 0.0150 & 0.2068 \\
       & 0    & 0.2156 & 0.0015 & -0.0190 & 0.0048 & 0.2030 \\
& $\ln(1.3)$  & 0.2156 & 0.0013 & -0.0148 & 0.0001 & 0.2022 \\
& $\ln 2$     & 0.2156 & 0.0011 & -0.0103 & -0.0032 & 0.2031 \\
\hline
\multirow{4}{15mm}{LO+bLO, $s_0=6.862$} &
$\ln(0.75)$   & 0.1458 & 0.0229 & 0.0238 & 0.0145 & 0.2070 \\
       & 0    & 0.1323 & 0.0308 & 0.0224 & 0.0175 & 0.2030 \\
& $\ln(1.3)$  & 0.1221 & 0.0354 & 0.0219 & 0.0175 & 0.1970 \\
& $\ln 2$     & 0.1085 & 0.0399 & 0.0224 & 0.0164 & 0.1872 \\ 
\end{tabular}
\end{ruledtabular}
\end{table}
The dependence of these results on the renormalization scale $\mu^2$ 
is graphically presented in Fig.~\ref{rtauvsRScl}.
We see that the leading-$\beta_0$-resummed (i.e., LB+bLB) 
evaluation method of 
Eq.~(\ref{rtLBbLBmanN}), in the depicted renormalization scale range 
$0.75 m_{\tau}^2 \leq |\mu^2| \leq 2 m_{\tau}^2$,
gives results for $r_{\tau}(\Delta S=0,m_q=0)$ that are 
significantly less renormalization scale dependent than those of the simpler
evaluation method of Eq.~(\ref{rtLObLOmanN}).

We can compare these results with the corresponding results
in perturbative QCD. We recall that the perturbative coupling $a_{\rm pt}(\mu^2)$
and our analytic coupling $\A_1(\mu^2)$ come together starting at renormalization scale 
$\mu^2=(3 m_c)^2$ upwards, and we have 
$a_{\rm pt}((3 m_c)^2) = \A_1((3 m_c)^2) \approx 0.07050$
in the renormalization scheme $\beta_2=\beta_3=\cdots =0$ [this value corresponds to
the value $a(M_Z^2,{\overline {\rm MS}}) \approx 0.119/\pi$, see Sec.~\ref{sec:mod};
and to $a(m_{\tau}^2,\beta_2=\beta_3=\cdots =0) = 0.3065/\pi$].
The perturbative QCD evaluation cannot use LB-resummation because Landau
poles appear in the LB-integral. Therefore, we can compare
only with the results of the LO+bLO method in perturbative QCD (using
the logarithmic derivatives $\ta_{\rm pt, n}$'s in the contour
integrals) and with the results of the usual perturbative ``power'' expansion
method (using powers of $a_{\rm pt}$ in the contour integral).
The results for $r_{\tau}$ are given in Table \ref{rtcomp},
for the two aforementioned methods in our analytic QCD model,
and for the LO+bLO and ``power'' expansion method in perturbative QCD,
at various values of the renormalization scale parameters ${\cal C}$.
\begin{table}
\caption{Values of $r_{\tau}$ evaluated by the truncated
analytic expansions Eq.~(\ref{rtLBbLBmanN}) [LB+bLB] and 
Eq.~(\ref{rtLObLOmanN}) [LO+bLO]
(for the values of the $s_0$ parameter $s_0=3.858$ and $6.862$,
respectively), as well as values obtained by perturbative evaluations
LO+bLO (involving $\ta_{\rm pt, n}$'s in the contour integral) 
and the truncated ``power'' expansion
(involving powers $a_{\rm pt}^n$'s in the contour integral), 
in the renormalization scheme $\beta_2=\beta_3 = \cdots =0$.
At renormalization scale $(3 m_c)$, analytic QCD and perturbative QCD are assumed to merge:
$a_{\rm pt}((3 m_c)^2) = \A_1((3 m_c)^2) \approx 0.07050$.
The results are truncated sums of four terms. 
The renormalization scale parameter ${\cal C}$ is 
varied from $\ln(3/4)$ to $\ln 2$.}
\label{rtcomp}  
\begin{ruledtabular}
\begin{tabular}{l|ll|ll}
${\cal C}$ & $r_{\tau}$: LB+bLB &  LO+bLO & pQCD LO+bLO & pQCD power exp.
\\
\hline
$\ln(0.75)$ & 0.2068 & 0.2070 & 0.1893 & 0.1856 \\ 
0         & 0.2030 & 0.2030 & 0.1873 & 0.1828 \\
$\ln(1.3)$  & 0.2022 & 0.1970 & 0.1850 & 0.1801 \\
$\ln2.$     & 0.2031 & 0.1872 & 0.1809 & 0.1753
\end{tabular}
\end{ruledtabular}
\end{table}
We see from Table \ref{rtcomp} that the variation
$|\Delta r_{\tau}|$ when the renormalization scale parameter ${\cal C}$ varies
between $\ln (0.75)$ and $\ln 2$ is for the four methods
46, 198, 84, 103, respectively. The facility of unambiguous
LB-resummation, which is possible only in analytic QCD models,
leads to reduced renormalization scale dependence of the result for $r_{\tau}$.
On the other hand, in perturbative QCD approaches the use of the logarithmic 
derivatives (pQCD LO+bLO) has the tendency to reduce the renormalization scale dependence
of the result for $r_{\tau}$ in comparison with the use of
the power expansion, something already noted in Ref.~\cite{CLMV}.\footnote{
In Ref.~\cite{CLMV}, perturbative QCD evaluation of the contour integral of
$r_{\tau}$ was performed in ${\overline {\rm MS}}$ renormalization scheme, 
and in that scheme the 
difference between the renormalization scale dependence of the two approaches
LO+bLO (there named: modified CIPT) and the power expansion approach
(CIPT) was found to be even greater than in the here presented
case of the renormalization scheme $\beta_2=\beta_3=\cdots=0$.}
We also note that in perturbative QCD, in order to reproduce the correct value
of $r_{\tau} \approx 0.203$, we need a larger value of 
$a_{\rm pt}$ that would correspond to $a_{\rm pt}(M_Z^2,{\overline {\rm MS}})
\approx 0.121/\pi$ (\cite{CLMV}); in our presented cases, we have
$a_{\rm pt}(M_Z^2,{\overline {\rm MS}}) \approx 0.119/\pi$, and therefore
perturbative QCD gives too low a value of $r_{\tau} \approx 0.18-0.19$. 

Another inclusive low energy QCD observable is,
for example, Bjorken polarized sum rule (BjPSR)
$d_{\rm Bj}(Q^2) = a(Q^2) + {\cal O}(a^2)$ at low $Q^2$. 
This is a spacelike quantity whose experimental
values are included in the last two lines of Table \ref{tabBj},
for three representative values of $Q^2$: $1.01, 1.71$ and $2.92 \ {\rm GeV}^2$.
These experimental data are based on the JLab CLAS EG1b (2006) 
measurements \cite{Deur2008}
of the $\Gamma_1^{\rm p-n}(Q^2)$ sum rule for spin-dependent proton and neutron
structure functions $g_1^{\rm p,n}$ \cite{Kataev:1994gd}.
The measured quantity $\Gamma_1^{\rm p-n}$ and the ``canonical'' 
BjPRS quantity $d_{\rm Bj}$ are related with each other in the following way:
\ba
\Gamma_1^{\rm p-n}(Q^2) &\equiv& \int_0^1 \; d x_{\rm Bj} \;
\left( g_1^{\rm p}(x_{\rm Bj},Q^2) -  g_1^{\rm n}(x_{\rm Bj},Q^2) \right) 
\label{Bjdef}
\\
&=&
\frac{g_A}{6} \left( 1 - d_{\rm Bj}(Q^2) \right) +
\sum_{j=2}^{\infty} \frac{ \mu_{2 j}^{\rm p-n} (Q^2) }{ (Q^2)^{j-1} } \ ,
\label{G1}
\ea
where $g_A = 1.267 \pm 0.004$ \cite{PDG2008} is the triplet axial charge,
$(1 - d_{\rm Bj})$ is the nonsinglet leading-twist
Wilson coefficient, while $\mu_{2 j}^{\rm p-n}/Q^{2 j -2}$ ($j \geq 2$)
are the higher-twist contributions. The measured JLab values of
$\Gamma_1^{\rm p-n}(Q^2)$, with the elastic contribution excluded, are
\cite{Deur2008}: $0.1236 \pm 0.0254$ for $Q^2 = 1.01 \ {\rm GeV}^2$;
$0.1605 \pm 0.0195$ for $Q^2 = 1.71 \ {\rm GeV}^2$;
$0.1789 \pm 0.0112$ for $Q^2 = 2.92 \ {\rm GeV}^2$.
The values given in the last two lines of Table \ref{tabBj} are 
obtained from these values by subtracting from the aforementioned
measured values the first higher-twist term $\mu_4^{\rm p-n}/Q^2$
with the value 
$\mu_4^{\rm p-n} \approx \mu_4^{\rm p-n}(Q=1{\rm GeV}) = -0.040 \pm 0.028$
obtained by a three-parameter perturbative QCD fit in Ref.~\cite{Deur2008}. The
central value ($-0.040$) was reconfirmed in 
Refs.~\cite{Pase1,Pase2} by a fit using the MA
\cite{ShS,MSS} approach. 
In addition, in Ref.~\cite{Pase1} it was shown, with the perturbative QCD 
and MA approach, that the the exclusion of the elastic contribution 
leads to strongly suppressed coefficients $\mu_4^{\rm p-n}$
at the higher-twist terms $\sim 1/(Q^2)^{j-1}$ 
with $j \geq 3$.
In the second line of experimental values in Table \ref{tabBj}, 
the uncertainties were split into the contribution coming from the 
uncertainty of the measured value of 
$\Gamma_1^{\rm p -n}$ and the one from the uncertainty of the
fitted value $\mu_4^{\rm p-n}$.

The first two coefficients $d_1$ and $d_2$ in the expansion of 
$d_{\rm Bj}$ were obtained in Refs.~\cite{LV} (given there in ${\overline {\rm MS}}$
renormalization scheme); the third coefficient $d_3$ is not known exactly, but
estimates are known, e.g., Ref.~\cite{KS}): in ${\overline {\rm MS}}$
scheme and at renormalization scale $\mu^2=Q^2$, it is $({\bar d}_{\rm Bj})_3 = 130.$;
we will use this value. 
The characteristic function 
$F_{\cal D}^{\cal {E}}(t)$ for the leading-$\beta_0$ resummation
for BjPSR was calculated and used in Ref.~\cite{CV1} 
(on the basis of the known \cite{Broadhurst:1993ru}
leading-$\beta_0$ parts of coefficients), 
and was presented in Ref.~\cite{CV2}.
This allows us to apply the evaluation methods of
Eq.~(\ref{LBbLBmanN}) and Eq.~(\ref{manN})
in our analytic QCD model (in $\beta_2=\beta_3=\cdots=0$ renormalization scheme) for
$d_{\rm Bj}(Q^2)$. The results of these two methods, at
three different low squared momenta $Q^2$
($Q^2 = 1.01, 1.71, 2.92 \ {\rm GeV}^2$) are presented in
Table \ref{tabBj}.
\begin{table}
\caption{Bjorken polarized sum rule (BjPSR) results $d_{\rm Bj}(Q^2)$
evaluated with the truncated ``modified analytic expansions'' 
Eq.~(\ref{LBbLBmanN}) [LB+bLB, with $s_0=3.858$]
and Eq.~(\ref{manN}) [LO+bLO, with $s_0=3.858, 6.862$], for $N=4$ and $N=3$.
The renormalization scale parameter is taken ${\cal C}=0$  ($\mu^2 = Q^2 \exp({\cal C})$).
The number of active quark flavors is $n_f=3$.
In brackets the variation of the result is given when the
parameter ${\cal C}$ increases from zero to $\ln 2$,
and when the renormalization scheme changes (at ${\cal C}=0$) from 
$\beta_2=\beta_3=\cdots=0$ to ${\overline {\rm MS}}$, respectively.
The experimentally measured values are given in the last two lines 
(see the text for details).
} 
\label{tabBj} 
\begin{ruledtabular}
\begin{tabular}{ll|lll}
$s_0$ & method & $d_{\rm Bj}(Q^2):\  Q^2=1.01 \ {\rm GeV}^2$ & $Q^2=1.71 \ {\rm GeV}^2$ &
$Q^2=2.92 \ {\rm GeV}^2$
\\
\hline 
\multirow{2}{8mm}{$3.858$} & LB+bLB, $N=4$ & $0.1973 \ [+2.1\%,-10.0\%]$ & 
$0.1755 \ [+7.3\%,-7.8\%]$ & $0.1595 \ [+7.2\%,-4.3\%]$
\\
 & LB+bLB, $N=3$ & $0.2290 \ [+6.1\%,-5.3\%]$ & 
$0.2066 \ [+5.5\%,-3.1\%]$ & $0.1827 \ [+4.3\%,-0.8\%]$
\\
\hline
\multirow{2}{8mm}{$3.858$} & LO+bLO, $N=4$ & $0.2774 \ [-1.3\%,-1.3\%]$ & 
$0.2234 \ [-4.2\%,-0.6\%]$ & $0.1779 \ [-4.6\%,+0.0\%]$
\\
 & LO+bLO, $N=3$ & $0.2597 \ [-8.7\%,+1.5\%]$ & 
$0.2061 \ [-8.7\% ,+2.9\%]$ & $0.1650 \ [-7.5\%,+3.3\%]$
\\
\hline
 \multirow{2}{8mm}{$6.862$} & LO+bLO, $N=4$ & $0.2103 \ [+9.1\%,-0.7\%]$ & 
$0.1926 \ [+3.0\%,-1.1\%]$ & $0.1668 \ [-1.0\%,-0.7\%]$
\\
  & LO+bLO, $N=3$ & $0.2184 \ [+0.0\%,-2.2\%]$ & 
$0.1898 \ [-4.0\% ,-0.5\%]$ & $0.1598 \ [-5.5\%,+1.1\%]$
\\
\hline
 & \multirow{2}{8mm}{exp.} & $0.23 \pm 0.18$ & $0.13 \pm 0.12$ & $0.09 \pm 0.07$ 
\\
 & & $0.23 \pm 0.12 \pm 0.13$ & $0.13 \pm 0.09 \pm 0.08$ & $0.09 \pm 0.05 \pm 0.05$
\end{tabular}
\end{ruledtabular}
\end{table}
We present in Table \ref{tabBj} the results both in the case
when the ${\rm N}^3 {\rm LB}$ (and ${\rm N}^3 {\rm LO}$) terms
of $\sim \A_4$ are included in $d_{\rm Bj}(Q^2)$ ($N=4$ case), 
and when they are not included ($N=3$ case).
Further, variations of the results under the change of renormalization scale
and scheme are also given. The renormalization scale was varied 
from the original $\mu^2=Q^2$
(${\cal C}=0$) to $\mu^2=2 Q^2$ (${\cal C} = \ln 2$).
The renormalization scheme was varied from the original 
scheme $\beta_2=\beta_3=\cdots =0$
to the ${\overline {\rm MS}}$ scheme $\beta_2=10.0599$ and $\beta_3=47.2281$
($n_f=3$). 

The change of the renormalization scheme was performed in the following way
(cf.~Ref.~\cite{CV2}). The dependence of couplings 
on the renormalization scheme
parameters $c_2 \equiv \beta_2/\beta_0$ and  $c_3 \equiv \beta_3/\beta_0$
is governed by the partial differential equations (pDFs) that are obtained
from the corresponding pDFs of perturbative QCD under the
analytization rule (\ref{anrule})
\ba
\frac{\partial \A_1(Q^2;c_2,c_3)}{\partial c_2} & = &
\frac{1}{2} \frac{ \partial^2 \A_1}{\partial x^2} +
\frac{5}{12} c_1 \frac{ \partial^3 \A_1}{\partial x^3} \ ,
\label{pDFc2A1}
\\
\frac{\partial \tA_2(Q^2;c_2,c_3)}{\partial c_2} & = &
\frac{1}{2} \frac{ \partial^2 \tA_2}{\partial x^2} \ ,
\label{pDFc2tA2}
\\
\frac{\partial \A_1(Q^2;c_2,c_3)}{\partial c_3} & = &
-\frac{1}{12} \frac{ \partial^3 \A_1}{\partial x^3} \ ,
\label{pDFc3A1}
\ea
where $x \equiv \beta_0 \ln(Q^2/\Lambda^2)$. The left-hand sides of these
pDFs are truncated, i.e., terms of $\sim \A_5$ ($\sim A_1^5$) are
ignored, because the truncated series for $d_{\rm Bj}(Q^2)$ is
known only up to $\sim \A_4$ (if the aforementioned
estimated value of $(d_{\rm Bj})_3$ is used).

The results in Table \ref{tabBj} show that the values of $d_{\rm Bj}(Q^2)$
obtained with the (LB+bLB) method of Eq.~(\ref{LBbLBmanN})
with $N=4$ are worse than those 
obtained with $N=3$, since the renormalization scale and scheme 
dependence is in general
stronger in the $N=4$ case. This has to do with the numerical
behavior of the $d_{\rm Bj}(Q^2)$ series in the approach 
of Eq.~(\ref{LBbLBmanN}) in this
model, because the fourth term (${\rm N}^3 {\rm LB}$, $\sim \tA_4$) is 
comparable or even larger than the third term 
(${\rm N}^2 {\rm LB}$, $\sim \tA_3$)
in this approach. E.g., for $Q^2=1.71 \ {\rm GeV}^2$, the series is
(when ${\cal C}=0$ and renormalization scheme $\beta_2=\cdots =0$):
$d_{\rm Bj}(Q^2) \approx 0.248 - 0.014 - 0.027 - 0.031 - \cdots$.
We conclude that the leading-$\beta_0$ resummed approach
of Eq.~(\ref{LBbLBmanN}) is not working well for $d_{\rm Bj}(Q^2)$ at low $Q^2$,
i.e., that the leading-$\beta_0$ terms are numerically not
representative (of at least some) of the perturbative coefficients 
$({\widetilde d}_{\rm Bj})_n$ ($n=2,3,\ldots$). The latter fact is shown 
in Table \ref{tabLBcoef}.
\begin{table}
\caption{The coefficients $({\widetilde d}_{\rm Adl})_n$ and
$({\widetilde d}_{\rm Bj})_n$ for $n=1,2,3$, and their leading-$\beta_0$
(LB) counterparts, in two renormalization schemes: $\beta_2=\beta_3=\cdots=0$,
and in ${\overline {\rm MS}}$ renormalization scheme. The renormalization scale parameter is ${\cal C}=0$, 
and $n_f=3$.}
\label{tabLBcoef} 
\begin{ruledtabular}
\begin{tabular}{c|c|ll|ll|ll}
Quantity & RSch & ${\widetilde d}_1$ & ${\widetilde d}_1^{\rm (LB)}$ &
${\widetilde d}_2$ &${\widetilde d}_2^{\rm (LB)}$ &${\widetilde d}_3$ &
${\widetilde d}_3^{\rm (LB)}$
\\
\hline
\multirow{2}{15mm}{Adler} & $\beta_2=\beta_3=\cdots=0$ & 
1.640 & 1.556 & 7.93 & 15.71 & 39.00 & 24.83
\\
  & ${\overline {\rm MS}}$ &
1.640 & 1.556 & 3.46 & 15.71 & 26.38 & 24.83
\\
\hline
\multirow{2}{15mm}{BjPSR} & $\beta_2=\cdots=0$ & 
3.583 & 4.5 & 18.32 & 32.34 & 91.13 & 255.23
\\
  & ${\overline {\rm MS}}$ &
3.583 & 4.5 & 13.84 & 32.34 & 52.45 & 255.23
\end{tabular}
\end{ruledtabular}
\end{table}
In that Table, the perturbative coefficients  $({\widetilde d}_{\rm Adl})_n$ 
and their leading-$\beta_0$ (LB) counterparts are also given; 
these coefficients were relevant in the evaluation of $r_{\tau}$.
Note that the leading-$\beta_0$ coefficients are renormalization 
scheme independent, they depend
only on the renormalization scale.
Comparing the coefficients $({\widetilde d}_{\rm Bj})_3$ and
$({\widetilde d}_{\rm Adl})_3$ with their leading-$\beta_0$ counterparts,
we can understand why the approach of Eq.~(\ref{LBbLBmanN})
with $N=4$ is expected to
work better in the evaluation of $r_{\tau}$ than in the
evaluation of $d_{\rm Bj}(Q^2)$. We recall that the ${\rm N}^3 {\rm LB}$
term in this evaluation of a spacelike observable
${\cal D}(Q^2)$ is $T_3 \tA_4(e^{\cal C} Q^2)$ where
$T_3 = ({\widetilde d}_4 - {\widetilde d}_4^{\rm (LB)})$,
cf.~Eqs.~(\ref{LBbLBmanN})-(\ref{Tn}).

On the other hand, Table \ref{tabBj} shows that the simpler 
approach of Eq.~(\ref{manN}) (LO+bLO), 
in our model gives results for $d_{\rm Bj}(Q^2)$
that in general get more stable under the 
renormalization scale and scheme variations 
when the number of terms increases from
$N=3$ to $N=4$, for $Q^2 \geq 1.7 \ {\rm GeV}^2$. The case of very
low scale $Q^2=1.01 \ {\rm GeV}^2$ is an exception, and probably has
to do with the fact that our analytic QCD model is not very reliable at
such low energies. We note that for $s_0=6.862$, where this approach is 
also used, the threshold masses are relatively high:
$M_1= \sqrt{s_1} \Lambda = 0.612$ GeV and $M_0 = \sqrt{s_0} \Lambda = 1.275$ GeV.

However, in order to see whether we have any better convergence
behavior in the evaluation of $d_{\rm Bj}(Q^2)$ than in the perturbative QCD,
we should compare with the perturbative QCD evaluation of $d_{\rm Bj}(Q^2)$.
The perturbative coupling and our analytic coupling merge starting at renormalization scale
$\mu^2=(3 m_c)^2$ upwards, where we have 
$a_{\rm pt}((3 m_c)^2) = \A_1((3 m_c)^2) \approx 0.07050$,
in the renormalization scheme $\beta_2=\beta_3=\cdots =0$ (this value corresponding to
the value $a(M_Z^2,{\overline {\rm MS}}) \approx 0.119/\pi$, see Sec.~\ref{sec:mod}).
In Table \ref{Bjcomp} we present the values of the evaluated expansion
terms for $d_{\rm Bj}(Q^2)$ at various $Q^2$, in the approach of
Eq.~(\ref{manN}) in our model 
(i.e., in terms of $\tA_n$'s), in the analogous approach in perturbative QCD 
(i.e., in terms of $\ta_{{\rm pt},n}$'s), and in
the usual power expansion approach in perturbative QCD (in powers
of $a_{\rm pt}$), using the renormalization scale $\mu^2=Q^2$ (i.e., ${\cal C}=0$), all in the same renormalization scheme
$\beta_2=\beta_3=\cdots =0$. 
\begin{table}
\caption{The four terms in truncated analytic expansions
(\ref{manN}) of $d_{\rm Bj}(Q^2)$ in our analytic QCD model with $s_0=6.862,3.858$; 
and the terms in the corresponding perturbative expansion in $\ta_{{\rm pt},n}$'s
(``pQCD LO+bLO''); 
and the terms in the usual power expansion,
for $Q^2 = 1.01, 1.71$ and $2.92 \ {\rm GeV}^2$.
The renormalization scale is chosen as $\mu^2=Q^2$; renormalization scheme is: $\beta_2=\beta_3=\cdots=0$;
in parentheses, the corresponding values in ${\overline {\rm MS}}$ renormalization scheme
are given.} 
\label{Bjcomp}  
\begin{ruledtabular}
\begin{tabular}{l|r|rrrr|r}
$Q^2$ [${\rm GeV}^2$] & method & $d_{\rm Bj}(Q^2)$: LO & NLO & 
${\rm N}^2{\rm LO}$ & ${\rm N}^3{\rm LO}$ & sum 
\\
\hline
\multirow{4}{8mm}{$1.01$}  
 & LO+bLO ($s_0=6.862$) & 0.1329 (0.1384) & 0.0621 (0.0575) & 0.0234 (0.0177) & -0.0081 (-0.0047) & 0.2103 (0.2089) \\
 & LO+bLO ($s_0=3.858$) & 0.1367 (0.1471) & 0.0746 (0.0798) & 0.0485 (0.0366) & 0.0176 (0.0102) & 0.2774 (0.2738) \\
 & pQCD LO+bLO  & 0.1398 (0.1603) & 0.0874 (0.1368) & 0.0857 (0.1587) &  0.0861 (0.2071) & 0.3989 (0.6628) \\
 & pQCD power exp.  & 0.1398 (0.1603) & 0.0700 (0.0920) & 0.0674 (0.0832) &  0.0658 (0.0858) & 0.3430 (0.4213) \\
\hline
\multirow{4}{8mm}{$1.71$}                            
 & LO+bLO ($s_0=6.862$) & 0.1142 (0.1198) & 0.0507 (0.0502) & 0.0249 (0.0188) &  0.0028 (0.0016) & 0.1926 (0.1905) \\
 & LO+bLO ($s_0=3.858$) & 0.1154 (0.1229) & 0.0551 (0.0622) & 0.0356 (0.0269) &  0.0173 (0.0099) & 0.2234 (0.2219) \\
 & pQCD LO+bLO  & 0.1162 (0.1266) & 0.0584 (0.0769) & 0.0454 (0.0596) &  0.0360 (0.0506) & 0.2559 (0.3136) \\
 & pQCD power exp.  & 0.1162 (0.1266) & 0.0484 (0.0574) & 0.0387 (0.0410) &  0.0314 (0.0333) & 0.2347 (0.2583) \\
\hline
\multirow{4}{8mm}{$2.92$}                            
 & LO+bLO ($s_0=6.862$) & 0.0991 (0.1040) & 0.0397 (0.0418) & 0.0209 (0.0158) &  0.0070 (0.0040) & 0.1668 (0.1656) \\
 & LO+bLO ($s_0=3.858$) & 0.0994 (0.1045) & 0.0411 (0.0474) & 0.0245 (0.0186) &  0.0128 (0.0074) & 0.1779 (0.1779) \\
 & pQCD LO+bLO  & 0.0996 (0.1057) & 0.0418 (0.0505) & 0.0269 (0.0298) &  0.0176 (0.0188) & 0.1860 (0.2047) \\
 & pQCD power exp.  & 0.0996 (0.1057) & 0.0355 (0.0400) & 0.0244 (0.0239) & 0.0170 (0.0162) & 0.1765 (0.1857) \\
\end{tabular}
\end{ruledtabular}
\end{table}
In addition, the corresponding values in the ${\overline {\rm MS}}$
renormalization scheme are also given there (in parentheses). We see that the
evaluated expansions in perturbative QCD for $d_{\rm Bj}(Q^2)$ at low momenta
$Q^2 \approx 1$-$2 \ {\rm GeV}^2$ behave much worse than the
expansion Eq.~(\ref{manN}) in our model; in fact, the third and the fourth
terms are roughly the same in these perturbative expansions (in $\ta_{{\rm pt},n}$'s;
and in powers) in the renormalization scheme 
$\beta_2=\beta_3=\cdots = 0$. Consequently, the
expansions in perturbative QCD turn out to be very unreliable
at such low values of $Q^2$.
The situation in these perturbative QCD expansions becomes even worse in the 
${\overline {\rm MS}}$ renormalization scheme; this has to
do primarily with the fact that the offending Landau cut
goes quite far into the positive regime in this scheme; the branching point
(Landau pole) is at $Q_b \approx 0.388$ GeV and $0.627$ GeV
in the two schemes, respectively.

The experimentally measured values 
are (Ref.\cite{Deur2008}):
$0.23 \pm 0.18$ for $Q^2=1.01 \ {\rm GeV}^2$; $0.13 \pm 0.12$ for
$Q^2=1.71 \ {\rm GeV}^2$; $0.09 \pm 0.07$ for $Q^2 =2.92 \ {\rm GeV}^2$;
and are given also in Table \ref{tabBj}
Comparing the results of Tables \ref{tabBj} and \ref{Bjcomp} with these values,
we see that the results of both methods, Eqs.~(\ref{manN}) and
(\ref{LBbLBmanN}), in the presented analytic QCD model,
with $s_0=3.858$ and $s_0=6.862$, lie in general above the central 
experimental values, but in general
within the large $1 \sigma$ uncertainties of the experimental values.
On the other hand, the perturbative QCD results for $Q^2 = 1.01 \ {\rm GeV}^2$ 
and $Q^2 = 1.71 \ {\rm GeV}^2$ are significantly higher than those
of the analytic QCD model. Some of the perturbative results
lie outside the $1 \sigma$ interval of experimental values, and
they show in general significantly worse convergence properties
than the analytic QCD results. 
Furthermore, the experimental
results indicate the tendency to lower values when $Q^2$ increases, 
and this is also the case of all the results in Table \ref{tabBj}.
In the case of evaluation of Bjorken polarized sum rule in the
present analytic QCD model,
the method of Eq.~(\ref{manN}), i.e., with no leading-$\beta_0$
resummation, should be regarded as the more reliable one.
This is in contrast with
the results for $r_{\tau}$ where we saw that the method of
Eq.~(\ref{rtLBbLBmanN}), which involves the leading-$\beta_0$ resummation,
is in our analytic model more reliable and less renormalization scale dependent.

\section{Conclusions}
\label{sec:concl}

We presented a simple analytic QCD model which has initially three dimensionless
parameters, and the scale parameter $\Lambda$. 
The model is obtained by
parametrizing, in a specific renormalization scheme, the unknown behavior 
of the discontinuity function of the coupling in the IR regime by one delta function. 
The number of parameters in the model reduces from four to one by 
imposing the requirement that the analytic coupling $\A_1(Q^2)$ differ from 
the perturbative coupling $a_{\rm pt}(Q^2)$ at high $Q^2$ only by a small amount 
$\sim (\Lambda/Q^2)^3$. Therefore, the model merges with perturbative QCD at high 
energies $|Q^2| > 10^1 \ {\rm GeV}^2$ to a high degree of accuracy, and reproduces 
all the values of the high-energy QCD observables  (with $|Q^2| > 10^1 \ {\rm GeV}^2$)
just like the perturbative QCD does.

The remaining free dimensionless parameter $s_0$ is then adjusted so that the 
model reproduces
the well measured strangeless semihadronic tau decay ratio $r_{\tau}$ -- in the case when
the leading-$\beta_0$ resummation is performed [Eq.~(\ref{rtLBbLBmanN});
$s_0=3.858$] in the evaluation, and in the case when it is not performed 
[Eq.~(\ref{rtLObLOmanN}); $s_0=6.862$]. 
The evaluated results for the
tau decay ratio in the approach of Eq.~(\ref{rtLBbLBmanN}) turn out to be 
quite stable under the variation of the renormalization scale and show
good convergence, not quite so the approach of Eq.~(\ref{rtLObLOmanN}).
On the other hand, the evaluated values of the Bjorken polarized sum rule 
$d_{\rm Bj}(Q^2)$ at low momentum transfer $Q^2 < 3 \ {\rm GeV}^2$
in the presented model behave reasonably well 
under the variation of the renormalization scale and scheme, and show a
reasonable good convergence, if no leading-$\beta_0$ resummation is performed; 
i.e., in the case of $d_{\rm Bj}(Q^2)$  the approach of Eq.~(\ref{manN}) 
gives better results than the approach of Eq.~(\ref{LBbLBmanN}). The
perturbative QCD evaluations of $d_{\rm Bj}(Q^2)$ at such low values
of $Q^2$ turn out to be very unreliable.

It remains an outstanding problem how to perform in a numerically efficient
way the change of the renormalization scheme for complex values of squared
momentum transfer $Q^2$ in analytic QCD models in general, and in the presented
model in particular. Solution of this problem would shed light on
the degree of stability of the evaluated tau decay ratio $r_{\tau}$ under the
renormalization scheme variation.

Another interesting problem would be to parametrize the unknown
behavior of the discontinuity function of the coupling in the IR regime
by two or more delta functions. This would allow us to fulfill 
the condition of merging the model with perturbative QCD 
(at $|Q^2| > 10^1 \ {\rm GeV}^2$) 
to an even higher degree of accuracy $\sim (\Lambda/Q^2)^5$,
and would thus allow us to 
apply and interpret the Operator product expansion technique 
in such analytic QCD models in an analogous way as in perturbative QCD.
  
\begin{acknowledgments}
\noindent
This work was supported in part by FONDECYT (Chile) Grant No.~1095196 
(C.C., G.C., O.E.), FONDECYT (Chile) Grant No.~1100348 (O.E.),
Rings Project No.~ACT119 (G.C.), and DAAD (H.E.M.).
\end{acknowledgments}

\end{document}